\documentclass[11pt,a4paper]{article} 
\usepackage{jcappub}
\usepackage[T1]{fontenc}
\usepackage{feynmf}

\title{\boldmath CMB and Random Flights: temperature and polarization in position space}

\author[a,b]{Paulo H. F. Reimberg}
\author[a]{L. Raul Abramo}

\affiliation[a]{Instituto de F\'isica, Universidade de S\~ao Paulo, 
CP 66318, 05314-970, S\~ao Paulo, Brazil}
\affiliation[b]{Institut de Physique Th\'eorique, CEA-Saclay, 91191
Gif-sur-Yvette, France}

\emailAdd{reimberg@fma.if.usp.br}
\emailAdd{abramo@fma.if.usp.br}

\abstract{The fluctuations in the temperature and polarization of the cosmic microwave 
background are described by a hierarchy of Boltzmann equations. In its integral 
form, this Boltzmann hierarchy can be converted from the usual Fourier-space base
into a position-space and causal description. 
We show that probability densities for random flights play a key role in this description. 
The integral system can be treated as a perturbative series in the number of 
steps of the random flights, and the properties of random flight probabilities impose 
constraints on the domains of dependence. 
We show that, as a result of these domains, a Fourier-Bessel decomposition can 
be employed in order to calculate the random flight probability densities. 
We also illustrate how the H-theorem applies to the cosmic microwave background:
by using analytical formulae for the asymptotic limits of these probability densities, 
we show that, as the photon distribution approaches a state of equilibrium, 
both the temperature anisotropies and the net polarization must vanish.}

\begin{document}

%\pacs{98.80.-k,98.70.Vc,05.40.Fb}

\maketitle
\flushbottom

%%%%%%%%%%%%%%%%%%%%%%%%%%%%%%%%%%%%%%%%%%%%%%
\section{Introduction}

The cosmic microwave background (CMB) has been a prolific source of
theoretical \cite{ruth} and experimental 
\cite{Smoot:1992td,2012arXiv1212.5225B,2013arXiv1303.5062P} 
results over the last couple of decades.
Because it relies on low-energy interactions well inside the linear regime,
the CMB now occupies a central part in our understanding of Cosmology, 
providing crucial information about the initial state of the Universe and
about many processes that distorted it since the surface of last scattering.

A fundamental description of the temperature and polarization of the CMB 
starts with the basic interaction 
dynamics (free propagation and scatterings with matter sources), and 
is fully realized through the Einstein-Boltzmann hierarchy of equations.
We shall here describe the system in a backward fashion: starting with 
the integral version of Boltzmann's equations, we will uncover details 
about the physical system described by that set of equations. Our 
procedure will be directed by the underlying structure of those integral 
equations, in particular by the way in which the hierarchy couples the equations. 
By removing and then reintroducing the source terms we will be able to reveal 
some hidden details about the microscopic dynamics of the physical processes 
that lead to the properties of the CMB which are observed today. 
To be clearer, we shall demonstrate that the Boltzmann equations preserve, 
in a very fundamental way, the fact that the fluctuations in the temperature and 
polarization of the CMB were imprinted through Thomson scatterings of low 
energy photons and free electrons during the recombination. Consequently, the 
hierarchy can be rewritten in terms of a series expansion over the number of 
scatterings that photons suffered before decoupling from matter. In this series 
expansion, besides the probability for each interaction, the number of 
scatterings appears as a key ingredient, since it determines the probability 
density that mediates the way in which the source terms imprint their 
contribution into the final signal.

The argument which leads to these conclusions will take us through a treatment
of the CMB in position space and, in many senses, the present work is a 
further development of \cite{cmb_box}. When performing the conversion from 
Fourier space to position space, it is unavoidable that a certain class of integrals 
over products of spherical Bessel functions appear in the description. 
The proper understanding and interpretation of these integrals constitute the 
key barrier that we should overcome in the description of the system in position 
space. As we shall see, these integrals codify the memory of the fundamental 
scattering dynamics. In fact, we shall demonstrate that these integrals 
constitute a generalization of random flight probability densities. 

Random flights are a classical problem in mathematical physics 
\cite{watson}, with many applications in physics and astronomy 
\cite{chandrasekhar_random}. The problem was first proposed in the 
beginning of the 20$^{\rm th}$ century in context of the study of bird migrations. 
Lord Rayleigh, soon after, applied the same ideas in acoustics. 
Further contributions on this subject are described in ref. \cite{dutka}. 
In very simple terms, a random flight (in a $D$-dimensional space) is 
the trajectory performed by a body which moves at constant speed and
changes its direction of motion into another random direction at Poisson-distributed 
time intervals. If the movement has a fixed origin, we may ask, based on the length 
of the intermediate displacements, as well as on the number of displacements, 
what is the probability for the moving body to reach a distance $r$ from the origin.

The precise form of the visibility function also plays a fundamental role
in the description of the CMB in terms of random flights, since it models the 
probability of scatterings during the recombination --- which changes as a 
function of time only. The spatial independence of the visibility function and the elasticity of 
Thomson scattering (which are, of course, already codified in the usual 
Boltzmann hierarchy) are the fundamental ingredients which lead to a 
random-flight-based description of the CMB. 

This paper is organized as follows: after introducing the Boltzmann hierarchy 
in the integral form in section \ref{boltzmann_eq_sec}, we temporarily decouple 
the evolution of the temperature from the polarization, and show the 
consequences of this simplification for the expression of the temperature and 
polarization in section \ref{uncouple_temp}. We then show, in 
section \ref{random_flight_sec}, how the 
family of integrals over spherical Bessel functions that appears in that 
description is related to random flight probability densities functions. 
After clarifying the structure of this dynamical system under the approximation 
of uncoupled temperature evolution, we go back to the original form of the 
hierarchy in section \ref{full_coupled}, and show that the formalism 
introduced in the previous Sections can be extended to treat the general case. 
We present, in section \ref{fourier_bessel_sec}, a computational tool based on 
Fourier-Bessel expansions which is suitable for computing the probability 
functions appearing in our description. Finally, in section \ref{H_theorem_sec} 
we use asymptotic arguments to show that high order terms should not 
contribute to the observables, which is expected from Boltzmann's H-theorem.

%%%%%%%%%%%%%%%%%%%%%%%%%%%%%%%%%%%%%%%%%%%%
\section{Boltzmann's equations}
\label{boltzmann_eq_sec}
The hierarchy of Boltzmann's equations describing CMB temperature fluctuations and polarization can be written in terms of a set of coupled integral equations --- see, e.g., \cite{Seljak:1996is}. Let the temperature anisotropies observed at position $\bold{x}_o$
and at (conformal) time $\eta_o$ along the direction $\hat{\bold{o}}$ be given in 
terms of its momenta as:
\begin{equation}
\label{temperatura}
\Theta(\bold{x}_o, \eta_o, \hat{\bold{o}}) = \int \frac{d^3 \bold{k}}{(2 \pi)^{3/2}}
\, \mathrm{e}^{i \bold{k} \cdot \bold{x}_o}
\, 4 \pi\sum_{l m} 
i^l  
\, \Theta_l(\bold{k}, \eta_o) 
\, \mathrm{Y}_{lm}^*(\hat{\bold{k}}) 
\, \mathrm{Y}_{lm}(\hat{\bold{o}})  \; .
\end{equation}
Similarly, the polarization, in terms of the usual Stokes parameters $Q$, $U$ and $I$, 
is decomposed as:
\begin{eqnarray}
\label{polarizacao}
\frac{Q+iU}{4I} (\bold{x}_o, \eta_o, \hat{\bold{o}})& = & 
\int \frac{d^3 \bold{k}}{(2 \pi)^{3/2}}  
\, \mathrm{e}^{i \bold{k} \cdot \bold{x}_o}
\, 4 \pi \sum_{lm} \, i^l \, \alpha_{l} (\bold{k}, \eta_o) 
\, \mathrm{Y}_{l m}^* (\hat{\bold{k}})  \,
\phantom{ }_2\mathrm{Y}_{lm} (\hat{\bold{o}}) \; ,
\end{eqnarray}
where $\phantom{ }_2\mathrm{Y}_{lm}$ are the spin-weighted spherical harmonics \cite{straumann_cosmology}.

The momenta of the CMB temperature and polarization are then given by the
integral equations:
\begin{eqnarray}
\label{temperatura_coef}
\Theta_l(\bold{k}, \eta_o) & = & 
\int_0^{\eta_o} \, d\eta \, \mathrm{e}^{-\mu(\eta)} 
\Bigg\{ 
\mu'(\eta) 
\Bigg[  
\theta_{SW}(\bold{k}, \eta) \, j_l (k \Delta \eta_0) 
- k \, V_b(\bold{k}, \eta)  \, j_l' (k \Delta \eta_0) 
\nonumber \\ &  & +
\frac{1}{2} 
\Big[ \Theta_2(\bold{k}, \eta) - \sqrt{6} \alpha_2 (\bold{k}, \eta) 
\Big] 
\left[ \frac{3}{2}
j_l''(k\Delta \eta_0) + \frac12
j_l(k \Delta \eta_0) \right] \Bigg] 
\nonumber\\ 
&  & +  (\Psi' + \Phi')(\bold{k}, \eta) \, j_l (k \Delta \eta_0 ) \Bigg\} \; , 
\end{eqnarray}
and
\begin{eqnarray}
\label{polarizacao_coef}
\alpha_{l} (\bold{k}, \eta_o) 
=
-\frac{3}{2} \sqrt{\frac{(l+2)!}{(l-2)!}} \int_0^{\eta_o}  d\eta \,
\mu'(\eta) \mathrm{e}^{-\mu(\eta)} 
\, \frac{1}{2}\Big[ \Theta_2(\bold{k}, \eta) - \sqrt{6} \alpha_2 (\bold{k}, \eta) \Big] 
\frac{j_l(k \Delta \eta_0 )}{(k \Delta \eta_0)^2} \, .
\end{eqnarray}
In eqs. \eqref{temperatura_coef} -- \eqref{polarizacao_coef} a prime denotes
a derivative with respect to conformal time $\eta$, 
the optical depth to Thomson scattering is $\mu(\eta)$,
and we have defined the interval $\Delta \eta_0 = \eta_o-\eta$.
It is sometimes convenient to define the ubiquitous source term 
$P(\bold{k}, \eta)=\frac{1}{2} \left[ \Theta_2(\bold{k}, \eta) - \sqrt{6} \alpha_2 (\bold{k}, \eta) \right]$. The system above is closed once the perturbed Einstein equations are used to
determine how the linear scalar cosmological perturbations 
$\theta_{SW}$, $V_b$, $\Phi$ and $\Psi$ evolve with time. 
However, both the precise nature of the perturbed Einstein equations, or of the 
initial conditions that are used to evolve those equations, are irrelevant for our results.

Equations \eqref{temperatura_coef} -- \eqref{polarizacao_coef} show that the
primary sources of temperature fluctuations are the Sachs-Wolfe term, 
$\theta_{SW}$, the baryon velocity, $V_b$, and 
the gravitational potentials $\Phi$ and $\Psi$ (also known as the Bardeen potentials
--- we work in the conformal-Newtonian gauge). 
The primary source of the polarization of the CMB, on the other hand, 
is the quadrupole of the temperature fluctuations. The integral equations
then couple all the momenta of temperature and polarization, mediated by the 
visibility function $g(\eta) = \mu'(\eta) \mathrm{e}^{-\mu(\eta)}$.

Henceforth we will take $\bold{x}_o = 0$, i. e., the observer is taken to be 
at the origin of the coordinate system employed for the description of the problem.

%%%%%%%%%%%%%%%%%%%%%%%%%%%%%%%%%%%%%%%%%%%%
\section{Uncoupling the temperature evolution}
\label{uncouple_temp}
We can decouple eqs. \eqref{temperatura_coef} -- \eqref{polarizacao_coef} by 
neglecting the term $\alpha_2$ in eq. \eqref{temperatura_coef}. 
This truncation represents the approximation whereby deviations from the equilibrium 
temperature are described by the Sachs-Wolfe (SW) and integrated Sachs-Wolfe 
(ISW) effects. Within this approximation, then eq. \eqref{temperatura_coef} becomes:
\begin{eqnarray}
\label{temperatura_coef_simple}
\Theta_l (\bold{k}, \eta_o) & = & \int_0^{\eta_o} 
d\eta \, \Big\{ g(\eta) \left[ 
\theta_{SW}(\bold{k}, \eta) \, j_l (k \Delta \eta_0 ) 
-k \, V_b(\bold{k}, \eta)  \, j_l' (k \Delta \eta_0)  \right] 
\nonumber\\ &  & +
\left. \mathrm{e}^{-\mu(\eta)} (\Psi' + \Phi')(\bold{k}, \eta) \, j_l (k \Delta \eta_0 ) \right\} 
\; .
\end{eqnarray}
Neglecting polarization as a source term for the temperature anisotropies
is in fact a very good approximation, and the reason for this underlies 
the argument presented in this paper. The visibility function $g(\eta)$
should be regarded as the probability per unit (conformal) time that 
photons will scatter with some free electron --- and, in fact, $g(\eta)$ is 
defined in such a way that this probability is normalized, 
$\int_0^\infty d\eta \, g(\eta) = \int_0^\infty d\mu \, e^{-\mu} = 1$. 
This means that each time a factor of
the visibility function intermediates a source term, that source term is
damped by a factor $\epsilon$, with $0 < \epsilon < 1$.
Since the lowest-order polarization term has at least one factor of
the visibility function, it contributes as a source term to the temperature
with two factors of the visibility function. Hence, the SW and
ISW effects dominate the intensity of the signal, and 
eq. \eqref{temperatura_coef_simple} accounts for the largest 
contribution to the temperature anisotropies.

\subsection{CMB temperature in position space}

Let's now take the lowest-order contribution to the temperature anisotropies, 
and express it in terms of position space. 
Expressing eq. \eqref{temperatura} as:
\begin{equation}
\label{Theta_decomp}
\Theta(\eta_o, \hat{\bold{o}}) = \sum_{lm} \theta_{lm}^{(0)}(\eta_o)
\mathrm{Y}_{lm}(\hat{\bold{o}}) \, ,
\end{equation}
the coefficients $\theta_{lm}^{(0)}(\eta_o)$ are given by:
\begin{eqnarray}
\label{c_lm0_passo1}
\theta_{lm}^{(0)}(\eta_o) & = & 
2 \, i^l  \int_0^{\eta_o} d \eta  \int \frac{dk}{(2\pi)^{1/2}} 
\, k^2 
\int d^2 \hat{\bold{k}}  \, 
\mathrm{e}^{-\mu(\eta)} 
\nonumber \\ 
& & \times 
\left\{ \mu'(\eta) 
\left[  \theta_{SW}(\bold{k}, \eta) 
- V_b(\bold{k}, \eta)  
\frac{ \partial}{\partial \eta} \right] + (\Psi' + \Phi')(\bold{k}, \eta) \right\}
\mathrm{Y}_{lm}^*(\hat{\bold{k}}) \,  j_l (k \Delta\eta_o) \, 
\end{eqnarray}
We shall then define the primary source term operator as:
\begin{equation}
\label{S_lm}
S_{lm}(k, \eta) = \int d^2 \hat{\bold{k}}  \, \mathrm{e}^{-\mu(\eta)} 
\left\{ \mu'(\eta) 
\left[  \theta_{SW}(\bold{k}, \eta)  -  V_b(\bold{k}, \eta)  
\frac{ \partial}{\partial \eta} \right] + (\Psi' + \Phi')(\bold{k}, \eta) \right\}
\, \mathrm{Y}_{lm}^*(\hat{\bold{k}}) \, ,
\end{equation}
where we stress the fact that we have included the optical depth to
Thomson scattering in the definition of the source term.
In terms of eq. \eqref{S_lm}, eq. \eqref{c_lm0_passo1} can be written as:
\begin{equation}
\theta_{lm}^{(0)}  =  2 \, i^l \int_0^{\eta_o} d \eta \int \frac{dk}{(2\pi)^{1/2}} 
\, k^2 \, S_{lm}(k, \eta) \, j_l (k \Delta\eta_o ) \, .
\end{equation}
In order to obtain a position-space description, we will express
the coefficients $S_{lm}(k, \eta)$ in terms of their 
counterparts in position space, by means of a Hankel transform:
\begin{equation}
\label{hankel}
S_{lm}(k, \eta) = 2 \, (-i)^l \int \frac{dX}{(2 \pi)^{1/2}} \, X^2 \, S_{lm}(X, \eta) \, j_l(k\, X)  \; .
\end{equation}
Using now the orthogonality relation for spherical Bessel functions,
\begin{displaymath}
\int dk \, k^2 \, j_L(ak) \, j_L(bk) = \frac{\pi}{2} \, \frac{b^L}{a^{L+2}} \, \delta(a-b) \; ,
\end{displaymath}
we obtain:
\begin{equation}
\label{c_lm_0}
\theta_{lm}^{(0)}  = \int_0^{\eta_o} d\eta \, S_{lm}( X=\Delta\eta_0 , \eta)  \; .
\end{equation}
Eq. \eqref{c_lm_0} has a straightforward interpretation: it states that, in order 
for a source term at time $\eta$ contribute to the CMB signal at time $\eta_o$, 
that source must be located at the spherical shells of radius 
$\Delta \eta_0 = \eta_o-\eta$ centered at the observer. The set
of those spherical shells is a hypersurface which corresponds, of course, 
to the past light cone of the observer on $\{ {\bf{x}}_o, \eta_o \}$. 
Since the visibility function is highly peaked at the time of decoupling,
the primary source term contributes the most to the signal near the epochs
when $z(\eta) \simeq 1100$.

\subsection{CMB polarization in position space}

We shall now decompose the polarization as:
\begin{equation}
\label{qiu_soma_a}
\frac{Q+iU}{4I} (\eta_o, \hat{\bold{o}}) = \sum_{lm} \pi_{lm} (\eta_o)  
\, \phantom{  }_2\mathrm{Y}_{lm} (\hat{\bold{o}}) \; ,
\end{equation}
with the aim of determining the coefficients $\pi_{lm} (\eta_o) $. 
The source terms in eq. \eqref{polarizacao_coef} are 
$\Theta_2$ and $\alpha_2$, which are built iteratively from an initial temperature quadrupole.
We can, therefore, organize the iterative solution as a series into powers of the visibility function.
As a first step in the iterative solution, for example,
$\alpha_2$ (which is of higher order in the visibility function) will not be 
taken into account --- only the temperature quadrupole will contribute to 
generate polarization at this order. 
The first iteration is, therefore:
\begin{eqnarray}
\label{a_lm_1_step1}
\pi_{lm}^{(1)} (\eta_o) & = & - \frac{3}{4} \, 2 \, i^l \, \sqrt{\frac{(l+2)!}{(l-2)!}}
\int_0^{\eta_o} d \eta_1 \, g(\eta_1) \int_0^{\eta_1} d \eta \, 
\int \frac{d k}{(2 \pi)^{1/2}} \, k^2 \, S_{lm}(k, \eta) 
\nonumber \\ & & \times  
 j_2(k \Delta \eta_1 ) \, \frac{j_l(k \Delta \eta_0 )}{(k \Delta \eta_0 )^2} \; ,
\end{eqnarray}
where the time intervals are defined as
$\Delta \eta_1 = \eta_1-\eta$ and $\Delta \eta_0 = \eta_o - \eta_1$.
The source term $S_{lm}$ is the same as was defined in eq. \eqref{S_lm}. 
Using once again the Hankel transform of eq. \eqref{hankel} we can recast 
eq. \eqref{a_lm_1_step1} as:
\begin{eqnarray}
\label{alm_1_integral}
\pi_{lm}^{(1)} (\eta_o) & = & - \frac{3}{2\pi} \sqrt{\frac{(l+2)!}{(l-2)!}}
\int_0^{\eta_o} d \eta_1 \, g(\eta_1) \, \int_0^{\eta_1} 
d \eta \, \int_0^{\infty} dX \, X^2 \, S_{lm}(X, \eta) 
\nonumber \\ 
&  & \times 
\int_0^{\infty} dk \, k^2 \, j_l(kX) \, \frac{j_l(k\Delta\eta_o)}{(k \Delta\eta_o)^2}
\, j_2(k\Delta\eta_1) \, .
\end{eqnarray}
The interpretation of the expression above is the following: at time $\eta$ a source
term generates a temperature quadrupole. That quadrupole then generates, through
a scattering at time $\eta_1$, the polarization which is finally observed at time $\eta_o$.
As we shall see, the integral of the second line of eq. \eqref{alm_1_integral}
guarantees that the source term, at a distance $X$ from the origin, 
is located in the past lightcone of the observer, for all possible $\eta_1$ and $\eta$. The variable $X$ will also be upper-bounded, and therefore the upper limit in the integration over the source terms will be replaced by a finite value that, as we shall see, corresponds the radius of the observer's past lightcone.

The next step in the iterative solution is to take the $\pi_{2m}^{(1)}$ just computed 
and use it as a source term for the polarization itself --- this means that now the 
polarization piece of the source term in eq. \eqref{polarizacao_coef}, $\alpha_2$, is no 
longer assumed to vanish. This contribution, which we will call $\pi_{lm}^{(2)}$, is 
therefore given by:
\begin{eqnarray}
\pi_{lm}^{(2)}(\eta_o) & =& - \frac{3}{2} \, \sqrt{\frac{(l+2)!}{(l-2)!}} \,
\int_0^{\eta_o} d \eta_1 \, g(\eta_1) \,
\int \frac{d^3 \bold{k}}{(2 \pi)^{3/2}} \, 4 \pi \, i^l \, 
\Bigg[ - \frac{\sqrt{6}}{2} \, \alpha_{2} (\bold{k}, \eta_1)
\Bigg] 
\mathrm{Y}_{lm}^*(\hat{\bold{k}})  \,
\frac{j_l(k\Delta\eta_o)}{(k\Delta\eta_o)^2} 
\nonumber \\ 
& = &
-\frac{3}{2} \, \sqrt{\frac{(l+2)!}{(l-2)!}} \, 
\int_0^{\eta_o} d \eta_1 \, g(\eta_1) \, 4 \pi \, i^l \, \frac{1}{2 \pi} \,
\Bigg[ -\frac{\sqrt{6}}{2} \, \left( -\frac{3}{2} \right) \, \sqrt{24} \, 
\int_0^{\eta_1} \, d \eta_2 \, g(\eta_2) 
\nonumber \\ 
& & \times  
\frac{1}{2} \, \int \frac{dk}{(2 \pi)^{3/2}} \, k^2 \, \int d^2 \hat{\bold{k}}
\, \Theta_2(\bold{k},\eta_2) \, \mathrm{Y}_{lm}^*(\hat{\bold{k}}) \,
\frac{j_2(k\Delta\eta_1)}{(k\Delta\eta_1)^2} \Bigg] 
\frac{j_l(k\Delta\eta_o)}{(k\Delta\eta_o)^2}  \nonumber \; ,
\end{eqnarray}
where now the interval $\Delta\eta_1 = \eta_1-\eta_2$.
In terms of the primary sources in position space, after using eq. \eqref{c_lm_0}
we obtain:
\begin{eqnarray}
\label{pi2pos}
\pi_{lm}^{(2)}(\eta_o) 
& = & 
-\frac{3}{2\pi} \, \sqrt{\frac{(l+2)!}{(l-2)!}} 
\int_0^{\eta_o} d \eta_1 \, g(\eta_1) \;
\int_0^{\eta_2} \, d \eta_2 \, g(\eta_2) \, 
\int_0^{\eta_2} \, d \eta 
\nonumber \\ 
& & \times  
\int_0^{\infty}  dX  X^2  S_{lm}(X, \eta)
\Bigg[9 \int dk  k^2 
j_l(kX)  \frac{j_l(k\Delta\eta_o )}{(k\Delta\eta_o)^2} 
\frac{j_2(k\Delta\eta_1)}{(k\Delta\eta_1)^2} 
j_2(k\Delta\eta_2)   \Bigg] \, .
\end{eqnarray}
The term $\pi_{lm}^{(2)}(\eta_o)$ is weighted twice by the visibility function and corresponds, as we will show in details, to the contribution to the total polarization coming from photons that have Thomson scattered twice during the recombination. 

At this point it is important to clarify our notation. We will always count the photon scatterings backwards in time: the
time the photons are observed is always taken to be $\eta_o$; 
the last time that the photons scattered before being observed is $\eta_1$; 
and so on. By convention, we will always evaluate the primary source 
term $S_{lm}$ at the instant $\eta$, so the sequence of scatterings ends with $\eta$.
Hence if, as in the case described by eq. \eqref{pi2pos},
there are two scatterings between the generation of the signal at $\eta$
and its observation at $\eta_o$, then we have $\eta_o \geq \eta_1 \geq \eta_2 \geq \eta$, 
and the time intervals always express the differences between one time and the
previous one, so in that case $\Delta \eta_0 = \eta_o - \eta_1$, 
$\Delta \eta_1 = \eta_1 - \eta_2$, and 
$\Delta \eta_2 = \eta_2 - \eta$.

The same argument allows us to calculate the next order contribution, which
takes into account three intermediate scatterings:
\begin{eqnarray}
\label{a_lm_3}
\pi_{lm}^{(3)}(\eta_o) & = & 
-\frac{3}{2\pi} \, \sqrt{\frac{(l+2)!}{(l-2)!}} \, 
\int_0^{\eta_o} \, d \eta_1 \, g(\eta_1) \,
\int_0^{\eta_1} d \eta_2 \, g(\eta_2) \,
\int_0^{\eta_2} d \eta_3 \, g(\eta_3)
\nonumber \\ 
 & & \times 
\int_0^{\eta_3} d \eta \, \int_0^{\infty} \, dX \, X^2  \, S_{lm}(X, \eta)
\nonumber \\ 
& & \times 
9^2 \, \int  \, dk  \, k^2  \, j_l(kX)  \, 
\frac{j_l(k\Delta\eta_o)}{(k\Delta\eta_o)^2} \,
\frac{j_2(k\Delta\eta_1)}{(k\Delta\eta_1)^2} \,
\frac{j_2(k\Delta\eta_2)}{(k\Delta\eta_2 )^2} \,
j_2(k\Delta\eta_3)    \; .
\end{eqnarray}
The general term in the iterative series expansion with $n$ intermediate scatterings
can be written as:
\begin{eqnarray}
\label{a_lm_integrais}
\pi_{lm}^{(n)}(\eta_o) 
& = & 
-\frac{3}{2\pi} \, \sqrt{\frac{(l+2)!}{(l-2)!}} \, 
\int_0^{\eta_o} \, d \eta_1 \, g(\eta_1) 
\, \times \,
\frac{1}{(n-1)!} \int_0^{\eta_1} \, d \eta_2 \, \ldots \, d \eta_n \,
\mathrm{T} \{ g(\eta_2) \, \ldots \,  g(\eta_n) \} 
\nonumber \\ 
& & \times 
\int_0^{\eta_n} \, d \eta \, \int_0^{\infty} \, dX \, X^2  \, 
S_{lm}(X, \eta)
\nonumber \\ 
& & \times 
9^{(n-1)} \, \int \, dk \, k^2 \, 
j_l(kX) \,
\frac{j_l(k\Delta\eta_o)}{(k\Delta\eta_o)^2}   
\underbrace{
\frac{j_2(k\Delta\eta_1)}{(k\Delta\eta_1)^2} \,
\ldots \, 
\frac{j_2(k\Delta\eta_{n-1})}{(k\Delta\eta_{n-1})^2}
}_{(n-1) \, times}
 j_2(k\Delta\eta_n)   \, 
\end{eqnarray}
Here, $\mathrm{T}$ stands for the time-ordered product of the sub-intervals, 
whose purpose is to reproduce the chain of integrations mediated by 
visibility functions shown in, e.g., eq. \eqref{a_lm_3}. 

The coefficients $\pi_{lm}$ appearing in eq. \eqref{qiu_soma_a} can be expressed, therefore, as:
\begin{equation}
\pi_{lm}(\eta_o) = \sum_{n=0}^{\infty} \pi_{lm}^{(n)}(\eta_o) \, ,
\end{equation}
with $\pi_{lm}^{(n)}(\eta_o)$ given by eq. \eqref{a_lm_integrais}.

An important condition for the validity of this perturbative expansion of the
CMB temperature and polarization is that
all the terms in the expansion of eq. \eqref{qiu_soma_a}, with an arbitrary
number $n$ of intermediate scatterings, must be expressed in position space. 
However, this can only be true if the $k$ integrals over products of spherical 
Bessel functions appearing in eq. \eqref{a_lm_integrais} can be in fact performed,
and are well-behaved. In the next Section we will show that, in fact, these 
integrals are probability densities for random flights with $n$ steps, in a 
space of suitable dimensionality.

%%%%%%%%%%%%%%%%%%%%%%%%%%%%%%%%%%%%%%%%%%%%
\section{Random flights and the CMB}
\label{random_flight_sec}
%%%%%%%%%%%%%%%%%%%%%%%%%%%%%%%%%
From the previous Section --- specially from eq. \eqref{a_lm_integrais} --- it is evident 
that a complete treatment of the CMB in position space requires that some specific 
integrals of products of spherical Bessel functions should be computed. 
In order to fulfill this requirement, we will proceed in the following way: first, 
we will present a simplified version of those integrals, and we will show that they 
give rise to probability densities associated with random flights. Next, 
we will show how the integrals we have to solve can be expressed in terms of 
the random flight integrals. 
%%%%%%%%%%%%%%%%%%%%%%%%%%%%%%%%%

Let's recall the well-known identities satisfied by spherical Bessel functions:
\begin{displaymath}
z^{L+\frac32} J_{L+\frac12}(z) = \frac{d}{dz}
\left[ z^{L + \frac32} J_{L+\frac32}(z) \right] \; .
\end{displaymath}
Since the spherical Bessel functions are defined as:
\begin{displaymath}
j_n(z) = \sqrt{\frac{\pi}{2 \, z}} \, J_{n+\frac12}(z) \; ,
\end{displaymath}
we are able to write:

\begin{eqnarray}
\label{int_prob}
& & \int_0^{\infty} dk \, k^2 \, j_L(k \, r) 
\prod_{i=1}^{n-1}\frac{j_L(k \, r_i)}{(k \, r_i)^L} \, j_L(k \, r_n) 
= \frac{ \left(\frac{\pi}{2}\right)^{(n+1)/2}}{\left[ \Gamma \left( L+\frac32 \right) 
\right]^{n-1}} \frac{r_n^L}{r^{L+2}} 
\nonumber \\ 
& & \times \frac{d}{dr} \left\{  
\left[ \Gamma \left( L+\frac32 \right) \right]^{n-2} 
\int_0^{\infty} \, dk \, r \, \left( \frac{k \, r}{2} \right)^{L+\frac12} \,
J_{L+\frac32}(k \, r)
\prod_{i=1}^{n} \frac{J_{L+\frac12} (k \, r_i)}{(k \, r_i)^{L+\frac12}} \right\} \, .
\end{eqnarray}
The derivative of the second line in eq. \eqref{int_prob} is, in fact, the probability 
density associated with a random flight --- see, e.g., ref. \cite{watson}. 
This is the probability density that a particle which moves with a constant (and finite) 
speed, and which starts from a given position in space, will be at a distance $r$ from
the point of origin, after changing randomly directions $n$ times during its trajectory. 
The length of the intermediate steps are denoted by $r_i$, $i=1, \ldots, n$. 
The order $L$ of the spherical Bessel functions in these integrals is related to 
the dimensionality of the space where the flight takes place: namely, 
the dimension $D$ of that space is given by $D=2L+3$. 

Following the notation employed by ref. \cite{watson} we shall denote:
\begin{eqnarray}
\label{p_notation}
p_n(r; r_1, \ldots, r_n \, | \, 2L+3) & := &   
\frac{d}{dr} \Bigg\{ \left[ \Gamma \left( L+\frac32 \right) \right]^{n-1}
\nonumber\\ & & \times 
\int_0^{\infty} dk \, r \, \left( \frac{k r}{2} \right)^{L+\frac12} \, 
J_{L+\frac32}(kr) \,
\prod_{i=1}^{n} \frac{J_{L+\frac12} (k r_i)}{(k r_i)^{L+\frac12}} \Bigg\} \, .
\end{eqnarray}
However, the integral \eqref{int_prob} is still not what we need in order 
to solve the momentum integrals that appear in our iterations --- see, e.g.,
the Fourier integration of eq. \eqref{a_lm_integrais}. 

We will now extend the random flight integrals to include the scenario
that appears in the context of the CMB.
If $l \geq L$ (which is always the case in our iterative solutions), 
then the product of two spherical Bessel functions of order $l$ 
can be written in terms of a single spherical Bessel function of order $L$. This is
a consequence of Gegenbauer's relation \cite{watson, talman} and of the 
orthogonality of associated Legendre polynomials --- see \cite{cmb_box} for a derivation:
\begin{equation}
\label{produto_bessel}
j_{l} (k X) \, j_l (k R) = \frac{(-1)^L}{2} \, \int_{|X-R|}^{X+R} \, d r \, k^L \,
\left( \frac{X R}{r} \right)^{L-1} \, P_{l}^{-L} (\cos \alpha) \, \left( \sin \alpha \right)^L \, j_L(kr) \, 
\end{equation}
where $r$, $R$ and $X$ must form a triangle, with the angle $\alpha$ being given 
implicitly in terms of the relation $r^2=R^2 + X^2 -2RX \cos\alpha$, and 
$P_{l}^{-L} (\cos \alpha)$ is an associated Legendre polynomial. Applying eq. \eqref{produto_bessel} and eq. \eqref{int_prob} with $L=2$ we can recast the Fourier 
integral in eq. \eqref{a_lm_integrais} as:
\begin{eqnarray}
\label{pol_prob}
F_{l2}(\Delta \eta_0, X; \Delta \eta_1, \ldots, \Delta \eta_n)  
& := & \int_0^{\infty} dk \, k^2 \, j_l(kX) \, 
\frac{j_l(k \Delta \eta_0 )}{(k \Delta \eta_0 )^2} 
\prod_{i=1}^{n-1}
\frac{j_2(k \Delta \eta_i)}{(k \Delta \eta_i)^2}  \, j_2(k \Delta \eta_n) 
\nonumber \\
& = &  
\frac{1}{2} 
\frac{ \left(\frac{\pi}{2}\right)^{(n+1)/2}}
{\left[ \Gamma \left( \frac72 \right) \right]^{n-1}} 
\int d (\cos \alpha) \, \left( \frac{X \Delta \eta_n}{r^3} \right)^2 \,
P_l^{-2}(\cos \alpha) \, \sin^2 \alpha 
\nonumber \\  
& & \times   p_n(r; \Delta \eta_1, \ldots, \Delta \eta_n \, | \, 7) \; ,
\end{eqnarray}
where we used the notation introduced in eq. \eqref{p_notation}.

In eq. \eqref{pol_prob} we have introduced the function
$F_{l2} (\Delta \eta_0, X; \Delta \eta_1, \ldots, \Delta \eta_n)$, which denotes 
what we shall call \emph{extended random flight integrals}. 
In general, $F_{lL}$ is defined by eq. \eqref{F_lm_m}.

We should examine eq. \eqref{pol_prob} more carefully. 
As anticipated, the presence of the term $p_n(r ; \Delta \eta_1, \ldots, \Delta \eta_n  |  7)$ 
should not be surprising, due to the interpretation of a random flight process 
and its validity with respect to the physics of the recombination. 
The dimensionality ($D=2\times2+3 =7$) of the space associated with the 
random flight, however, is not yet fully understood. 
That dimension is determined by the order of the spherical Bessel functions 
which mediate the sources of anisotropies and the final CMB signal, but 
since only the quadrupole of the temperature fluctuation contributes to the 
polarization, the spherical Bessel function of order $2$ is the one that 
characterizes the random flight for the CMB. A possible explanation for this
dimension is that, after separating the angular dependence of the CMB from its 
radial and time dependence through the spherical harmonic decomposition, 
the light cone has only two dimensions left. Since the
multipole $L$ has $2L+1$ degrees of freedom, we end up with $2L+3$ dimensions
where our relevant variables can perform random flights.
However, a more refined argument to explain the dimension 7 is not yet known.

Looking back now at eq. \eqref{a_lm_integrais}, we recognize that for the 
$\pi_{lm}^{(n)}$ term in the polarization expansion, the probability 
density $p_n(r; \Delta \eta_1, \ldots, \Delta \eta_n \, | \, 7)$ appears clearly. 
The interpretation of an expansion in the number of interactions during 
recombination is therefore strengthened. We should remark, however, that 
explicit formulae for $p_n(r; \Delta \eta_1, \ldots, \Delta \eta_n \, | \, 7)$ 
are not known in the general case \cite{kolesnik_06,kolesnik_08}.

\begin{figure}
\begin{center}
\includegraphics[scale=0.5]{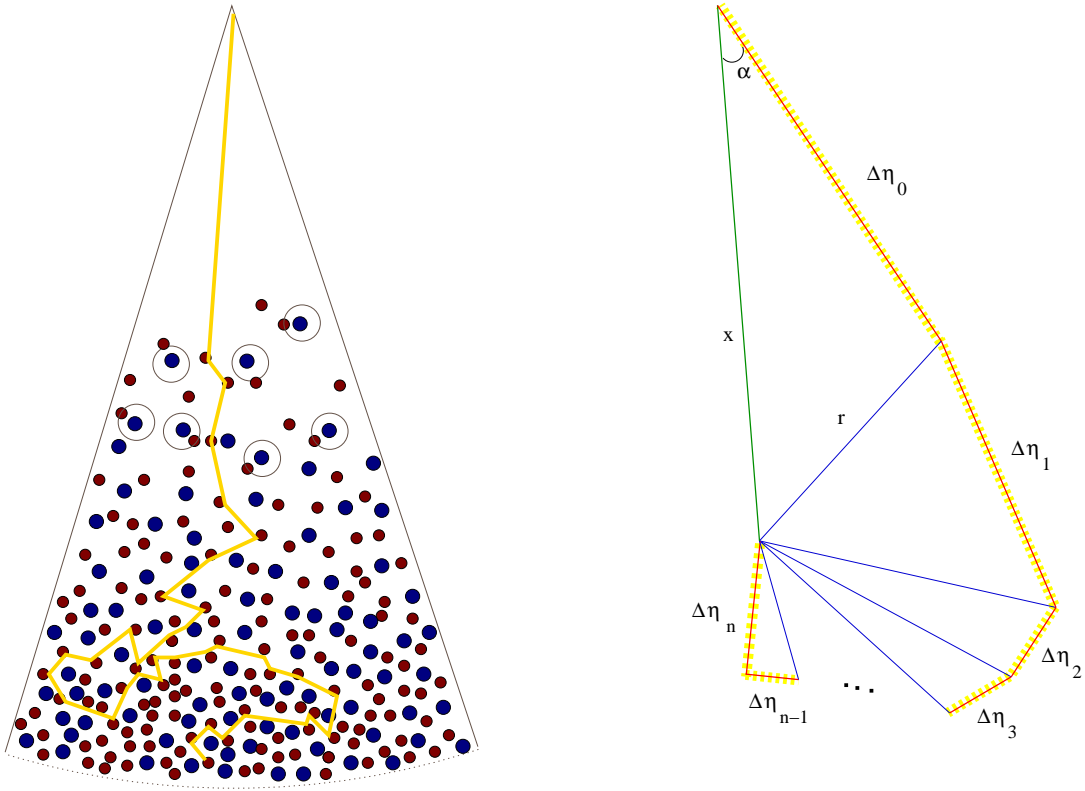}
\end{center}
\caption{\footnotesize Connection between scatterings of a photon during 
recombination (left) with a diagrammatic representation of the random flight (right). 
The steps $\Delta \eta_n, \ldots, \Delta \eta_1$ correspond to the lengths of 
the photon's trajectory between successive scatterings. 
$\Delta \eta_0$ corresponds to the propagation since the photon's last 
scattering during recombination, at time $\eta_1$, and the observation at 
time $\eta_o$. 
In both diagrams the observation takes place at the upper vertex.}
\label{juntos}
\end{figure}

As discussed above, the intervals $\Delta \eta_1, \ldots, \Delta \eta_n$ 
express the times elapsed between consecutive scatterings. 
All these subintervals are elements of a partition of the time 
interval $\eta_1 - \eta$, which is therefore the total time elapsed since the 
instant the photon leaves equilibrium with matter, at time $\eta$, until the instant 
$\eta_1$ when the photon has last scattered prior to its observation. 
This time interval represents, therefore, the effective duration of recombination for a 
given photon. The lengths of each subinterval are weighted by the visibility 
functions, and integrated in order to contemplate all possible histories for photons 
during recombination. 
The interval $\eta_o - \eta_1$ expresses the time elapsed since the photon scattered for the 
last time, before it is observed at the time $\eta_o$ (we are assuming that no 
further scatterings take place during this interval).
We can represent a photon's history by means of figure \ref{juntos}: on the left  
we show the photon's interactions prior to observation at the vertex of the cone, 
while on the right we show a diagrammatic representation of that history, 
with the relevant elements that appear in eq. \eqref{pol_prob}.

We call special attention to the diagram represented on the right of figure \ref{juntos}. 
This diagram represents the extended random flight performed by a photon during 
recombination. The steps $\Delta \eta_n$, \ldots, $\Delta \eta_1$ (going forward in time)
belong to a standard random flight, and describe the trajectory of a photon that has 
left equilibrium with matter at an instant $\eta$, then propagated freely for a distance 
(or a time interval) $\Delta \eta_n$, then Thomson-scattered with an electron at 
time $\eta_n$, then propagated freely for a distance $\Delta \eta_{n-1}$, and so on 
until the instant $\eta_1$, when it scattered for the last time. 
The standard random flight ends at the instant $\eta_1$. The photon, at that 
moment, is a radius $r$ away from the point where the flight started. 
The steps indicated by $\Delta \eta_0$ and $X$ do not belong to the standard 
random flight, but are present in the extended random flight, and are introduced 
through the spherical Bessel functions of different order in the $k$ 
integral of eq. \eqref{pol_prob}. This is necessary because those two steps are 
not associated with any movement between successive scatterings, but are in
fact associated with the distance from the observer to the origin of the photon, 
and to the end-point of the random flight. It should be strengthened that 
$\Delta \eta_0$, $X$ and $r$ are related by 
$r^2 = \Delta \eta_0^2 + X^2 - 2 \, \Delta \eta_0 \, X \, \cos \alpha$. 
Since $0 \leq r \leq \Delta \eta_1 + \Delta \eta_2 + \ldots + \Delta \eta_n$, it 
follows that 
$0 \leq X \leq \Delta \eta_0 + \Delta \eta_1 + \Delta \eta_2 + \ldots + \Delta \eta_n$, 
which then determines the domain of dependence of the problem.

%%%%%%%%%%%%%%%%%%%%%%%%%%%%%%%%%%%%%%%%%%%%
\section{The polarization in position space} 

We shall now go back to eq. \eqref{a_lm_integrais}. 
In terms of the extended random flight just introduced, the polarization
coefficient to $n^{th}$ order, $\pi_{lm}^{(n)}(\eta_o)$, 
can be written as:
\begin{eqnarray}
\label{a_lm_prob}
\pi_{lm}^{(n)} & = & 
-\frac{3}{4\pi} \, \sqrt{\frac{(l+2)!}{(l-2)!}} \, 
\int_0^{\eta_o} d \eta_1 \, g(\eta_1) \,
\frac{1}{(n-1)!} \int_0^{\eta_1} d \eta_2 \ldots \, d \eta_n \,
\mathrm{T} \{ g(\eta_2) \, \ldots \,  g(\eta_n) \} 
\nonumber \\ 
& & \times 
\frac{9^{n-1} \left(\frac{\pi}{2}\right)^{(n+1)/2}}
{\left[ \Gamma \left( \frac72 \right) \right]^{n-1}} 
\int_0^{\eta_n} d \eta \, \int_0^{\eta_o - \eta} \, dX \, X^2 \, 
S_{lm}(X, \eta) 
\nonumber \\ 
& & \times 
\int d (\cos \alpha) \, \left( \frac{X \Delta \eta_n}{r^3} \right)^2 \,
P_l^{-2}(\cos \alpha) \, \sin^2 \alpha \,  \, p_n(r; \Delta \eta_1, \ldots, \Delta \eta_n \, | \, 7) \; ,
\end{eqnarray}
where we have already used the aforementioned upper bound for the variable $X$. 

The equation \eqref{a_lm_prob} can be understood as the combination of three procedures:

\begin{itemize}

\item The integration over $\alpha$ corresponds to a 
marginalization over all possible paths composed of $n$ steps of 
lengths $\Delta \eta_1 + \ldots + \Delta \eta_n$ that have a net displacement 
$r$ determined by $X$ and $\Delta \eta_0$, as shown in figure \ref{marg_alpha}. 
This ``average over paths'' is a function of $X$, $\Delta \eta_0$, 
$\Delta \eta_1, \ldots, \Delta \eta_n$.

\begin{figure}
\begin{center}
\includegraphics[scale=0.7]{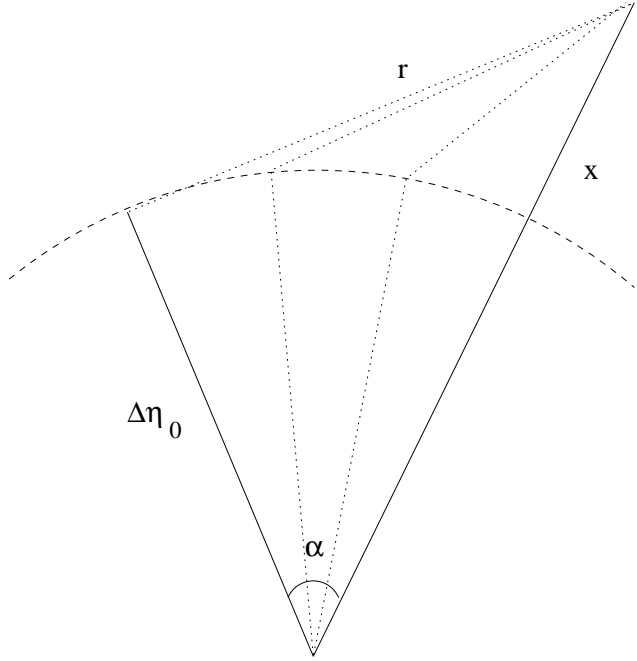}
\end{center}
\caption{\footnotesize Marginalization over all paths with $n$ steps, 
composed of the intermediate displacements $\Delta \eta_1, \ldots, \Delta \eta_n$, 
which lead to a fixed displacement $r$ with respect to the origin of the flight. 
The distance $r$ is determined by $X$ and $\Delta \eta_0$ for all possible angles 
$\alpha$.}
\label{marg_alpha}
\end{figure}

\item The contribution from the primary source term, 
$S_{lm}(X, \eta)$, is then mediated by this ``average over paths'' that was
just described, for all possible values of $X$. The maximum value that 
$X$ may reach is $\Delta \eta_0 + \Delta \eta_1 + \ldots + \Delta \eta_n = \eta_o - \eta$
, which is nothing but the radius of the observer's past light cone up to the time $\eta$. 
After computing the contribution of the source terms, we end up with an expression 
that is a function of $\Delta \eta_0 , \Delta \eta_1, \ldots, \Delta \eta_n$.

\item The last step is to let the intervals 
$\Delta \eta_0 , \Delta \eta_1, \ldots, \Delta \eta_n$ assume any values through 
the integrations, each one weighted by its corresponding factor of the visibility function
to take into account the probability that the photon will scatter at that instant of time.
This accomplishes the goal of accounting for the contribution from sources at all
distances, and over any possible number of intermediate steps of the extended random flights.
\end{itemize}

Adding the contributions from all $\pi_{lm}^{(n)}$ we obtain:
\begin{eqnarray}
\label{a_lm_total}
\pi_{lm} & = & -\frac{3}{4\pi} \sqrt{\frac{(l+2)!}{(l-2)!}} 
\int_0^{\eta_o} d \eta_1 \, g(\eta_1) 
\sum_{n=1}^{\infty}
\frac{1}{(n-1)!} \int_0^{\eta_1} d \eta_2 \ldots \,  d\eta_n \,
\mathrm{T} \{ g(\eta_2) \, \ldots \,   g(\eta_n) \} 
\nonumber\\ & & \times 
\frac{ 9^{n-1} \left(\frac{\pi}{2}\right)^{(n+1)/2}}
{\left[ \Gamma \left( \frac72 \right) \right]^{n-1}} 
\int_0^{\eta_n} d \eta \, \int_0^{\infty} dX \, X^2 \, S_{lm}(X, \eta)
\nonumber \\ & & \times 
\int d (\cos \alpha) \, \left( \frac{X \Delta \eta_n}{r^3} \right)^2
\, P_l^{-2}(\cos \alpha) \, \sin^2 \alpha \, \, 
p_n(r; \Delta \eta_1, \ldots, \Delta \eta_n \, | \, 7) \; .
\end{eqnarray}

In terms of the extended random flight integrals, eq. \eqref{pol_prob}, we can write:
\begin{eqnarray}
\pi_{lm} & = &
-\frac{3}{2\pi} \, 
\sqrt{\frac{(l+2)!}{(l-2)!}} \, 
\int_0^{\eta_o} d \eta_1 \, g(\eta_1) 
\, \sum_{n=1}^{\infty} \,
\frac{9^{n-1}}{(n-1)!} \, 
\int_0^{\eta_1} d \eta_2 \, \ldots \,  d\eta_n \mathrm{T} \{ \mu'(\eta') \mathrm{e}^{-\mu(\eta')} \ldots\,   g(\eta_n)  \} 
\nonumber \\ & & \times 
\int_0^{\eta_n} d \eta \, \int_0^{\infty} dX \, X^2  \, S_{lm}(X, \eta) \,
F_{l2} \left( \Delta \eta_0, X ; \Delta \eta_1, \ldots, \Delta \eta_n \right) \; .
\end{eqnarray}

%%%%%%%%%%%%%%%%%%%%%%%%%%%%%%%%%%%%%%%%%%%%
\section{Diagrams}
\label{diagrams_sec}
We are now able to introduce a diagrammatic representation for all the terms 
that appear in the series expansion of the temperature and polarization at $n$-th
order. The graphic elements are:
\begin{itemize}
\item Solid lines mean polarization, and dashed lines mean temperature;
\item The vertical lines to the right of the diagrams represent the 
observable being calculated (time runs upward);
\item The interception of two lines determine what sources are being considered
for a given observable.
\end{itemize}
The rules for constructing these diagrams will be detailed in the next Section,
but it is already easy to represent diagrammatically the contribution to
the CMB polarization computed in eq. \eqref{a_lm_total}:

%%%%%%%%%%%%%%
\begin{fmffile}{pol_decoup}
\begin{eqnarray}
\label{a_lm_tot_diag}
\parbox{20mm}{\begin{fmfgraph}(50, 110)
\fmfstraight
\fmfpen{thin}
\fmfright{i1,i2,i3}
\fmfleft{o1,o2,o3}
\fmf{phantom}{o1,o2,o3}
\fmf{vanilla}{i1,i2,i3}
\fmf{dashes}{i2,o2}
\fmfv{decor.shape=circle,decor.filled=empty,decor.size=.1w}{i2}
\end{fmfgraph}}
\quad  +  \quad
\parbox{20mm}{\begin{fmfgraph}(50, 110)
\fmfstraight
\fmfpen{thin}
\fmfright{i1,i2,i3,i4}
\fmfleft{o1,o2,o3,o4}
\fmf{phantom}{o1,o2,o3,o4}
\fmf{vanilla}{i1,i2,i3,i4}
\fmf{dashes}{i2,o2}
\fmf{dashes}{i3,o3}
\fmfv{decor.shape=circle,decor.filled=empty,decor.size=.1w}{i2}
\fmfv{decor.shape=circle,decor.filled=empty,decor.size=.1w}{i3}
\end{fmfgraph}}
\quad  +  \quad
\parbox{20mm}{\begin{fmfgraph}(50, 110)
\fmfstraight
\fmfpen{thin}
\fmfright{i1,i2,i3,i4,i5}
\fmfleft{o1,o2,o3,o4,o5}
\fmf{phantom}{o1,o2,o3,o4,o5}
\fmf{vanilla}{i1,i2,i3,i4,i5}
\fmf{dashes}{i2,o2}
\fmf{dashes}{i3,o3}
\fmf{dashes}{i4,o4}
\fmfv{decor.shape=circle,decor.filled=empty,decor.size=.1w}{i2}
\fmfv{decor.shape=circle,decor.filled=empty,decor.size=.1w}{i3}
\fmfv{decor.shape=circle,decor.filled=empty,decor.size=.1w}{i4}
\end{fmfgraph}}
\quad  + \quad \ldots \quad + \quad
\parbox{20mm}{\begin{fmfgraph}(50, 110)
\fmfstraight
\fmfpen{thin}
\fmfright{i1,i2,i3,i4,i5,i6,i7}
\fmfleft{o1,o2,o3,o4,o5,o6,o7}
\fmf{phantom}{o1,o2,o3,o4,o5,o6,o7}
\fmf{vanilla}{i1,i2,i3,i4}
\fmf{dots}{i4,i5}
\fmf{vanilla}{i5,i6,i7}
\fmf{dashes}{i2,o2}
\fmf{dashes}{i3,o3}
\fmf{dashes}{i6,o6}
\fmfv{decor.shape=circle,decor.filled=empty,decor.size=.1w}{i2}
\fmfv{decor.shape=circle,decor.filled=empty,decor.size=.1w}{i3}
\fmfv{decor.shape=circle,decor.filled=empty,decor.size=.1w}{i6}
\end{fmfgraph}}
\end{eqnarray}
\end{fmffile}
%%%%%%%%%%%%%%%%%%%%%%%%%%%%%%%%%%%%%%%%%%%%

\section{Full coupled temperature evolution}
\label{full_coupled}
We will now show that the formalism developed in the previous Sections is not
spoiled when all the contributions to the temperature fluctuations are included. 
Let's go back to eqs. \eqref{temperatura} and \eqref{temperatura_coef}, and write:
\begin{eqnarray}
\label{c_lm_passo1}
\theta_{lm} & = & 
2 \, i^l \int_0^{\eta_o} d \eta \int \frac{dk}{(2\pi)^{1/2}}  k^2 
\int d^2 \hat{\bold{k}}  \Bigg\{ 
g(\eta)  \Bigg[  \theta_{SW}(\bold{k}, \eta)  -  V_b (\bold{k}, \eta)  \frac{ \partial}{\partial \eta} \Bigg]   
\mathrm{e}^{-\mu(\eta)} (\Psi' + \Phi')(\bold{k}, \eta) 
\nonumber \\
& &  +  \frac{1}{2} \, g(\eta)  \,
P(\bold{k}, \eta) \, \left[ 1 + 3 \frac{\partial^2}{\partial (k\Delta\eta_o)^2} \right] 
\Bigg\} \, \mathrm{Y}_{lm}^*(\hat{\bold{k}}) \, j_l (k \Delta\eta_o ) \; . 
\nonumber
\end{eqnarray}
The first two terms in this equation have already been treated in Section III. 
We now proceed to analyzing the remaining term, i. e., that which contains 
the source 
$P(\bold{k}, \eta)=\frac{1}{2} \left[ \Theta_2(\bold{k}, \eta) - \sqrt{6} \, \alpha_2 (\bold{k}, \eta) \right]$. We write, therefore:
\begin{eqnarray}
\label{c_lm_passo2}
\theta_{lm} & = & \theta_{lm}^{(0)} + i^l \int_0^{\eta_o} d \eta \int 
\frac{dk}{(2\pi)^{1/2}} \, k^2 \int d^2 \hat{\bold{k}} \,
g(\eta) \, P(\bold{k}, \eta)  \mathrm{Y}_{lm}^*(\hat{\bold{k}})  
\, \left[ 1 + 3 \frac{\partial^2}{\partial (k\Delta\eta_o)^2} \right]  \,
j_l (k \Delta\eta_o)  
\nonumber\\ & =: &
\theta_{lm}^{(0)} + \sum_{n=1}^{\infty} \theta_{lm}^{(n)} \; ,
\end{eqnarray}
where we have defined the $\theta_{lm}^{(n)}$ in terms of the integral over sources.

\subsection{Temperature at first order}

We shall now calculate the lowest-order contribution described in eq. \eqref{c_lm_passo2} 
by taking $P(\bold{k}, \eta)$ to be $\frac{1}{2} \Theta_2(\bold{k}, \eta)$ --- i. e., only
the temperature quadrupole is taken as a source for the temperature fluctuations,
since the polarization (which is of higher order, since its source is the temperature 
quadrupole) is at least a second-order contribution to the temperature. 
The first order contribution to the temperature, $\theta_{lm}^{(1)}$, is therefore 
computed as:
\begin{eqnarray}
\label{c_lm1_passo1}
\theta_{lm}^{(1)} & = &  \frac{i^l}{2} \int_0^{\eta_o} d \eta_1 \, g(\eta_1) 
\int \frac{dk}{(2\pi)^{1/2}} \, k^2 \int d^2 \hat{\bold{k}} \,
\Theta_2(\bold{k}, \eta_1)  \, \mathrm{Y}_{lm}^*(\hat{\bold{k}})  
\left[ 1 + 3 \frac{\partial^2}{\partial (k\Delta\eta_o)^2} \right]  \, j_l (k \Delta\eta_o)  
\nonumber \\ 
& = &
\frac{i^l}{2} \int_0^{\eta_o} d \eta_1 \, g(\eta_1) 
\int \frac{dk}{(2\pi)^{1/2}} \, k^2
\int_0^{\eta_1} d \eta \, S_{lm}(k, \eta) \, j_2(k\Delta\eta_1) 
\left[ 1 + 3 \frac{\partial^2}{\partial (k\Delta\eta_o)^2} \right]  j_l (k \Delta\eta_o)
\nonumber\\ & = & 
\frac{1}{2\pi} \int_0^{\eta_o} d \eta_1 \, g(\eta_1)
\int_0^{\eta_1} d \eta \, \int_0^{\infty} dX \, X^2 \, 
S_{lm}(X, \eta)  
\nonumber\\ & & \times 
\int dk \, k^2 \, j_l(kX)  \, 
\left[ 1 + 3 \frac{\partial^2}{\partial (k\Delta\eta_o)^2} \right]  \, 
j_l (k \Delta\eta_o ) \, j_2(k\Delta\eta_1)
\end{eqnarray}
where we have once again transformed coefficients into the position-space description using Hankel transformation. We must now treat the integral in the last line of
the equation above:
\begin{displaymath}
I_{l2} = \int dk \, k^2 \, 
j_l(kX) \, \left[ 1 + 3 \frac{\partial^2}{\partial (k\Delta\eta_o)^2} \right]  
\, j_l (k \Delta\eta_o) \, j_2(k\Delta\eta_1) \, .
\end{displaymath}
Clearly, we can write:
\begin{eqnarray}
\label{c_lm_1_0}
I_{l2} & = & 
3 \left[ 
\frac{\partial^2}{\partial (\Delta\eta_o)^2} - \frac{1}{\Delta\eta_o}
\frac{\partial}{\partial (\Delta\eta_o)} 
\right] 
\left[ 
(\Delta\eta_o)^2 \int dk \, k^2 \, j_l(kX)  \, \frac{j_l(k\Delta\eta_o)}{(k\Delta\eta_o )^2}  
\, j_2(k\Delta\eta_1) 
\right] 
\nonumber\\ & & + \,
(\Delta\eta_o)^2 \int dk \, k^4 \, j_l(kX)  \,
\frac{j_l(k\Delta\eta_o)}{(k\Delta\eta_o)^2} 
\, j_2(k\Delta\eta_1) \; ,
\end{eqnarray}
which we can then express as:
\begin{eqnarray}
\label{c_lm_1}
I_{l2} =
3 \left[
\frac{\partial^2}{\partial (\Delta\eta_o)^2} - \frac{1}{\Delta\eta_o}
\frac{\partial}{\partial \Delta \eta_0} \right] 
\left[ (\Delta\eta_o)^2 
F_{l2} ( \Delta\eta_o, X ; \Delta\eta_1) \right] 
+  (\Delta\eta_o)^2  G_{l2}^{(2)} ( \Delta\eta_o, X; \Delta\eta_1) 
\end{eqnarray}
where the integral corresponding to $G_{l2}^{(2)}$ 
is related to the already introduced function $F_{l2}$ --- see eq. \eqref{pol_prob}. 
In fact, using an identity that is proved in the appendix, 
eq. \eqref{G_lm_k2_resultado}, we obtain:
\begin{eqnarray}
\theta_{lm}^{(1)} & = &
\frac{1}{2\pi} 
\int_0^{\eta_o} d \eta_1 \, g(\eta_1)
\int_0^{\eta_1} d \eta 
\int_0^{\infty} dX \, X^2 \, S_{lm}(X, \eta)  \,
\nonumber \\ 
& & \times 
\Bigg \{ 3 \left[\frac{\partial^2}{\partial (\Delta\eta_o)^2} - \frac{1}{\Delta\eta_o}
\frac{\partial}{\partial \Delta\eta_o} \right] 
\left[ (\Delta\eta_o)^2
F_{l2} (\Delta\eta_o, X ; \Delta\eta_1) \right] 
\nonumber\\ & & + 
\frac{1}{(\Delta\eta_o)^l \, X^{l+2}} \,
\frac{\partial}{\partial X} \,
\left( \frac{\partial}{\partial \Delta\eta_o} + \frac{2}{\Delta\eta_o} \right) 
\left[ (\Delta\eta_o \, X)^{l+2} F_{(l+1) \, 2} ( \Delta\eta_o, X; \Delta\eta_1) \right] 
\Bigg \}\, .
\end{eqnarray}
In conclusion, first-order corrections to the temperature 
can also be written in terms the probability densities of extended random flights.

%%%%%%%%%%%%%%%%%%%%%%%%%%%%%%%%%%%%%%%%%%%%%%
\subsection{Temperature at second order}

If we want to calculate further contributions to the temperature, then we must 
include the contributions to $\Theta$ that come from the temperature quadrupole, 
given in eq. \eqref{c_lm_1}, and those from $\alpha_2$, given in eq. 
\eqref{alm_1_integral}. Combining those two contributions leads to:
\begin{eqnarray}
\label{c_lm_2}
\theta_{lm}^{(2)} & = &
\frac{1}{4 \pi} 
\int_0^{\eta_o} d \eta_1 \, g(\eta_1)
\int_0^{\eta_1} d \eta_2 \, g(\eta_2) 
\int_0^{\eta_2} d \eta  
\int_0^{\infty} dX \, X^2 \, S_{lm}(X, \eta)  
\nonumber \\ 
& { } & \times
\int dk \, k^2 \, j_l(kX)  \, j_2(k\Delta\eta_2) 
\, \left[ 1 + 3 \frac{\partial^2}{\partial (k\Delta\eta_1)^2} \right]  
j_2 (k \Delta\eta_1) 
\left[ 1 + 3 \frac{\partial^2}{\partial (k\Delta\eta_o)^2} \right]  
j_l (k \Delta \eta_0) \, 
\nonumber \\ 
& { } & + \,  \frac{9}{\pi} 
\int_0^{\eta_o} d \eta_1 \, g(\eta_1) 
\int_0^{\eta_1} d \eta_2  \, g(\eta_2)
\int_0^{\eta_2} d \eta 
\int_0^{\infty} dX \, X^2 \, S_{lm}(X, \eta) 
\nonumber \\ 
& { } & \times
\int dk \, k^2 \, j_l(kX) \, j_2(k\Delta\eta_2) 
\frac{j_2(k\Delta\eta_1)}{(k\Delta\eta_1)^2} 
\left[ 1 + 3 \frac{\partial^2}{\partial (k\Delta \eta_0)^2} \right]  
j_l (k \Delta \eta_0) \; .
\end{eqnarray}
The first term comes from the contribution of 
$\Theta_2$ to $\theta_{lm}^{(2)}$, and the second term 
takes into account the contribution of $\alpha_2$ to $\theta_{lm}^{(2)}$.
These contributions can also be expressed in terms of the extended random 
flight integrals: terms with two derivatives can be expressed in 
terms of $F_{l2} (\Delta \eta_0, X; \Delta\eta_1, \Delta\eta_2)$; 
terms with one derivative can be expressed in terms of 
$F_{(l+1) \, 2} (\Delta \eta_0, X; \Delta\eta_1, \Delta\eta_2)$;
and terms with no derivatives can be expressed, as indicated in section \ref{high_power}, in terms of $F_{(l+2) \, 2} (\Delta \eta_0, X; \Delta\eta_1, \Delta\eta_2)$.

The complete series expansions for the temperature coefficients,
$\bar{\theta}_{lm}$, is represented diagrammatically, up to second order 
corrections, as:

%%%%%%%%%%%%%%%%%%%%%%%%%%%%%%%%%%%%%%%%%%%%%%%%
\begin{fmffile}{temp_2nd_ord}
\begin{eqnarray}
\bar{\theta}^{(2)}_{lm}=
\parbox{10mm}{\begin{fmfgraph}(10, 100)
\fmfstraight
\fmfpen{thin}
\fmftop{i1,i2,i3,i4}
\fmfbottom{o1,o2,o3,o4}
\fmf{dashes}{i4,o4}
\end{fmfgraph}}
 +  \quad
\parbox{20mm}{\begin{fmfgraph}(50, 100)
\fmfstraight
\fmfpen{thin}
\fmfright{i1,i2,i3}
\fmfleft{o1,o2,o3}
\fmf{phantom}{o1,o2,o3}
\fmf{dashes}{i1,i2,i3}
\fmf{dashes}{i2,o2}
\fmfv{decor.shape=circle,decor.filled=empty,decor.size=.1w}{i2}
\end{fmfgraph}}
\quad  +  \quad
\parbox{20mm}{\begin{fmfgraph}(50, 100)
\fmfstraight
\fmfpen{thin}
\fmfright{i1,i2,i3,i4}
\fmfleft{o1,o2,o3,o4}
\fmf{phantom}{o1,o2,o3,o4}
\fmf{dashes}{i1,i2,i3,i4}
\fmf{dashes}{i2,o2}
\fmf{dashes}{i3,o3}
\fmfv{decor.shape=circle,decor.filled=empty,decor.size=.1w}{i2}
\fmfv{decor.shape=circle,decor.filled=empty,decor.size=.1w}{i3}
\end{fmfgraph}}
\quad  +  \quad
\parbox{20mm}{\begin{fmfgraph}(50, 100)
\fmfstraight
\fmfpen{thin}
\fmfright{i1,i2,i3}
\fmfleft{o1,o2,o3}
\fmfbottom{b1,b2,b3}
\fmftop{t1,t2,t3}
\fmf{phantom}{o1,o2,o3}
\fmf{dashes}{i1,i2,i3}
\fmf{phantom}{t2,v}
\fmf{dashes}{v,b2}
\fmf{vanilla}{i2,o2}
\fmfv{decor.shape=circle,decor.filled=empty,decor.size=.1w}{i2}
\fmfv{decor.shape=circle,decor.filled=empty,decor.size=.1w}{v}
\end{fmfgraph}}
\end{eqnarray}
\end{fmffile}
The first diagram represents $\theta^{(0)}_{lm}$, the second,  
$\theta^{(1)}_{lm}$ --- see eq. \eqref{c_lm_1} --- and the two last 
diagrams represent  $\theta^{(2)}_{lm}$, as shown in eq. \eqref{c_lm_2}.
%%%%%%%%%%%%%%%%%%%%%%%%%

Even if we do not calculate here explicitly the contributions at all orders, the results 
presented in section \ref{high_power} show that all corrections can be 
expressed in terms of extended random flight integrals. 
In general, $\bar{\theta}_{lm}^{(n)}$ will be given by:
\begin{displaymath}
\bar{\theta}_{lm}^{(n)} = \sum_{q=0}^{n} 
\theta_{lmq}^{(n)} (\Delta \eta_0, \Delta \eta_1, \ldots, \Delta \eta_n, X) 
\, F_{(l+q)2} (\Delta \eta_0, X; \Delta \eta_1, \ldots, \Delta \eta_n) \; ,
\end{displaymath}
where we represent by $\theta_{lmq}^{(n)}$ the coefficients that appear in 
the expansion. As was seen explicitly in the case $n=1$, these coefficients 
are differential operators combined with powers of the subintervals of 
$(\eta_o - \eta)$ and of $X$. The complete expression for 
$\bar{\theta}_{lm}^{(n)}$ will also depend on knowledge of the expression for 
polarization up to the order $n-1$ --- which is the reason for leaving
the superscript $n$ explicit in $\theta_{lmq}^{(n)}$.
%%%%%%%%%%%%%%%%%%%%%%%%%%%%%%%%%%%%%%%%%%%%%%%%%%%%

\subsection{Polarization at second order}

The last step in our construction of the series expansions of the CMB in terms of extended
random flight probabilities is to treat the contribution of the temperature quadrupole 
to the polarization, allowing for the corrections to the temperature that come from higher 
order contributions of the source terms. We shall write, then:
\begin{eqnarray}
\bar{\pi}_{lm}^{(2)}  & = & -\frac{3}{2} \sqrt{\frac{(l+2)!}{(l-2)!}} 
\int_0^{\eta_o} d \eta_1 \, g(\eta_1) 
\int \frac{d^3 \bold{k}}{(2 \pi)^{3/2}} \, 4 \pi \, i^l \, 
\frac{1}{2} \, \Theta_2 (\bold{k}, \eta_1)  
\, \mathrm{Y}_{lm}^*(\hat{\bold{k}}) \,
\frac{j_l(k\Delta \eta_0)}{(k\Delta \eta_0)^2}
\nonumber \\ 
& = & 
-\frac{3}{2} \, i^l \, \sqrt{\frac{(l+2)!}{(l-2)!}} 
\int_0^{\eta_o} d \eta_1 \, g(\eta_1) 
\Bigg\{ \int_0^{\eta_1} d\eta \int \frac{dk}{(2\pi)^{1/2}} \, k^2 \, 
S_{lm}(k, \eta) \, j_2(k\Delta\eta_1) 
\nonumber \\ 
& & +  
\frac{1}{2} \int_0^{\eta_1} d \eta_2 \, g(\eta_2)
\int \frac{dk}{(2\pi)^{1/2}} \, k^2 \, 
\int d^2 \hat{\bold{k}} \, P(\bold{k}, \eta_2) \, 
\mathrm{Y}_{lm}^*(\hat{\bold{k}})  
\nonumber \\ 
& & \times 
\left[ 1 + 3 \frac{\partial^2}{\partial(k\Delta\eta_1)^2} \right]
j_2(k\Delta\eta_1) \Bigg\} \, \frac{j_l(k\Delta \eta_0)}{(k\Delta \eta_0)^2}  \; 
\end{eqnarray}
where the bar over $\bar{\pi}_{lm}^{(2)}$ means that we are taking all corrections 
up to second order.

At this point we can separate the first-order and the second-order terms, 
and focus on the contribution from $\Theta_2$ to $P$:
\begin{eqnarray}
\label{alm_sec}
\bar{\pi}_{lm}^{(2)}  & = & \pi_{lm}^{(1)} 
\, - \, \frac{3}{4} \, i^l \, \sqrt{\frac{(l+2)!}{(l-2)!}} 
\int_0^{\eta_o} d \eta_1 \, g(\eta_1) 
\int_0^{\eta_1} d \eta_2 \, g(\eta_2)
\int \frac{dk}{(2\pi)^{1/2}} \, k^2 \, 
\nonumber \\ 
& & \times 
\int d^2 \hat{\bold{k}} \, \Theta_2(\bold{k}, \eta_2) \, 
\mathrm{Y}_{lm}^*(\hat{\bold{k}}) \,
\left[ 1 + 3 \frac{\partial^2}{\partial(k\Delta\eta_1)^2} \right]
\, j_2(k\Delta\eta_1) 
\, \frac{j_l(k\Delta \eta_0)}{(k\Delta \eta_0)^2} 
\nonumber \\ 
& = &  
\pi_{lm}^{(1)} 
\, - \, \frac{3}{4} \, i^l \, 
\sqrt{\frac{(l+2)!}{(l-2)!}} 
\int_0^{\eta_o} d \eta_1 \, g(\eta_1) 
\int_0^{\eta_1} d \eta_2 \, g(\eta_2) 
\int \frac{dk}{(2\pi)^{1/2}} \, k^2 \, 
\nonumber \\ 
& & \times 
\int_0^{\eta_2} d \eta \, S_{lm}(k, \eta) \,
j_2(k\Delta\eta_2) \, \left[ 1 + 3 \frac{\partial^2}{\partial(k\Delta\eta_1)^2} \right] \, 
j_2(k\Delta\eta_1)  \,
\frac{j_l(k\Delta \eta_0)}{(k\Delta \eta_0)^2} 
\nonumber \\ 
& = &  \pi_{lm}^{(1)} \, - \, 
\frac{1}{2} \, \frac{3}{2 \pi} \, \sqrt{\frac{(l+2)!}{(l-2)!}}  
\int_0^{\eta_o} d \eta_1 \, g(\eta_1) 
\int_0^{\eta_1} d \eta_2 \, g(\eta_2) 
\int_0^{\eta_2} d \eta 
\int dX \, X^2 \, S_{lm} (X, \eta) \,
\nonumber \\ 
& & \times 
\Bigg\{
\int dk \, k^2 \, j_l(kX) \, \frac{j_l(k\Delta \eta_0)}{(k\Delta \eta_0)^2} \, 
\left[ 1 + 3 \frac{\partial^2}{\partial(k\Delta\eta_1)^2} \right] j_2(k\Delta\eta_1) 
\, j_2(k\Delta\eta_2) \Bigg\} \, .
\end{eqnarray}
The last integral of the equation above, between curly brackets, 
can be expressed, after using eq. \eqref{G_lm_k2_resultado}, as follows:
\begin{eqnarray}
\label{derivative_terms}
& &\int dk \, k^2 \, j_l(kX) \, 
\frac{j_l(k\Delta \eta_0)}{(k\Delta \eta_0)^2}  \,
\left[ 1 + 3 \frac{\partial^2}{\partial(k\Delta\eta_1)^2} \right] 
j_2(k\Delta\eta_1) \,
j_2(k\Delta\eta_2)
\nonumber \\ & & 
= \frac{(\Delta\eta_1)^2}{(\Delta \eta_0 \, X)^{l+2}} 
\, \frac{\partial}{\partial X}
\left( 
\frac{\partial}{\partial \Delta \eta_0} + 
\frac{2}{\Delta \eta_0} 
\right)
\Big[ (\Delta \eta_0 \, X)^{l+2} \,
F_{(l+1)2} (\Delta \eta_0, X; \Delta\eta_1, \Delta\eta_2) \Big] 
\nonumber \\ 
& & 
\, + \, 
3 \left( 
\frac{\partial^2}{\partial(\Delta\eta_1)^2} - \frac{1}{\Delta\eta_1} 
\frac{\partial}{\partial \Delta\eta_1} 
\right) 
\, \Big[ (\Delta\eta_1)^2 \, 
F_{l2} (\Delta \eta_0, X; \Delta\eta_1, \Delta\eta_2) \Big] \, .
\end{eqnarray}
As we can see, despite the somewhat cumbersome coefficients, the 
final result can be expressed, again, as a combination of 
the extended random flight functions 
$F_{l2} ( \Delta \eta_0, X; \Delta\eta_1, \Delta\eta_2)$ 
and $F_{(l+1)2} (\Delta \eta_0, X; \Delta\eta_1, \Delta\eta_2)$.

The last term in eq. \eqref{alm_sec} has the same order as $\pi_{lm}^{(2)}$. 
Therefore, to second order in temperature corrections, 
the polarization results from the following three diagrams:
%%%%%%%%%%%%%%
\begin{fmffile}{pol_2nd}
\begin{eqnarray}
\label{pol_sec_ord_diag}
\bar{\pi}^{(2)}_{lm} & = & 
\parbox{20mm}{\begin{fmfgraph}(50, 100)
\fmfstraight
\fmfpen{thin}
\fmfright{i1,i2,i3}
\fmfleft{o1,o2,o3}
\fmf{phantom}{o1,o2,o3}
\fmf{vanilla}{i1,i2,i3}
\fmf{dashes}{i2,o2}
\fmfv{decor.shape=circle,decor.filled=empty,decor.size=.1w}{i2}
\end{fmfgraph}}
\quad  +  \quad 
\parbox{20mm}{\begin{fmfgraph}(50, 100)
\fmfstraight
\fmfpen{thin}
\fmfright{i1,i2,i3,i4}
\fmfleft{o1,o2,o3,o4}
\fmf{phantom}{o1,o2,o3,o4}
\fmf{vanilla}{i1,i2,i3,i4}
\fmf{dashes}{i2,o2}
\fmf{dashes}{i3,o3}
\fmfv{decor.shape=circle,decor.filled=empty,decor.size=.1w}{i2}
\fmfv{decor.shape=circle,decor.filled=empty,decor.size=.1w}{i3}
\end{fmfgraph}}
\quad  +  \quad
\parbox{20mm}{\begin{fmfgraph}(50, 100)
\fmfstraight
\fmfpen{thin}
\fmfright{i1,i2,i3}
\fmfleft{o1,o2,o3}
\fmfbottom{b1,b2,b3}
\fmftop{t1,t2,t3}
\fmf{phantom}{o1,o2,o3}
\fmf{vanilla}{i1,i2,i3}
\fmf{phantom}{t2,v}
\fmf{dashes}{v,b2}
\fmf{dashes}{i2,o2}
\fmfv{decor.shape=circle,decor.filled=empty,decor.size=.1w}{i2}
\fmfv{decor.shape=circle,decor.filled=empty,decor.size=.1w}{v}
\end{fmfgraph}}
\end{eqnarray}
\end{fmffile}

As we showed above, this contribution can be expressed fully in terms of
extended random flight probabilities, therefore, up to second order the polarization
coefficients can be written as:
\begin{eqnarray}
\label{a_bar_2}
\bar{\pi}_{lm}^{(2)} 
&=& 
\pi_{lm\, 0}(\Delta \eta_0, \Delta\eta_1, \Delta \eta_2, X)
\, F_{l \, 2} (\Delta \eta_0, X; \Delta\eta_1, \Delta \eta_2) \, 
\\ \nonumber
& & +  \,
\pi_{lm\, 1}(\Delta \eta_0, \Delta\eta_1, \Delta\eta_2; X)
\, F_{(l+1) \, 2} (\Delta \eta_0, X; \Delta\eta_1, \Delta\eta_2 ) \, .
\end{eqnarray} 
The coefficients $\pi_{lmq}$ ($q=0,1$) are differential operators, and depend on the 
combinations of the subintervals to some power. The coefficient $\pi_{lm\, 0}$ takes into account the contributions from the $\pi_{lm}^{(1)}$ term appearing in eq. \eqref{alm_sec} and also the contribution coming from the second term in the right hand side of eq. \eqref{derivative_terms}. The coefficient $\pi_{lm\, 1}$ can be obtained by working out the first term in the right hand side of eq. \eqref{derivative_terms}.

If we examine eqs. \eqref{alm_sec} and \eqref{pol_sec_ord_diag}, 
it can be seen that $\bar{\pi}_{lm}^{(2)}$ is given by the sum of $\pi_{lm}^{(1)}$ and 
the contribution from the temperature corrected by the polarization. 
We can write eq. \eqref{pol_sec_ord_diag} in a more symmetric 
way if we note that the following diagrams are equivalent: 

%%%%%%%%%%%%%%%%%%%%%%%%%%%%%
\begin{fmffile}{pol_equiv}
\begin{eqnarray}
\label{equiv_diag}
\parbox{20mm}{\begin{fmfgraph}(50, 100)
\fmfstraight
\fmfpen{thin}
\fmfright{i1,i2,i3,i4}
\fmfleft{o1,o2,o3,o4}
\fmf{phantom}{o1,o2,o3,o4}
\fmf{vanilla}{i1,i2,i3,i4}
\fmf{dashes}{i2,o2}
\fmf{dashes}{i3,o3}
\fmfv{decor.shape=circle,decor.filled=empty,decor.size=.1w}{i2}
\fmfv{decor.shape=circle,decor.filled=empty,decor.size=.1w}{i3}
\end{fmfgraph}}
\qquad = \qquad
\parbox{20mm}{\begin{fmfgraph}(50, 100)
\fmfstraight
\fmfpen{thin}
\fmfright{i1,i2,i3}
\fmfleft{o1,o2,o3}
\fmfbottom{b1,b2,b3}
\fmftop{t1,t2,t3}
\fmf{phantom}{o1,o2,o3}
\fmf{vanilla}{i1,i2,i3}
\fmf{phantom}{t2,v}
\fmf{dashes}{v,b2}
\fmf{vanilla}{i2,o2}
\fmfv{decor.shape=circle,decor.filled=empty,decor.size=.1w}{i2}
\fmfv{decor.shape=circle,decor.filled=empty,decor.size=.1w}{v}
\end{fmfgraph}} 
\end{eqnarray}
\end{fmffile}
%%%%%%%%%%%%%%%%%%%%%%%%%%%%%
This equivalence is due to the fact that both diagrams have the 
same vertical lines at the right (we are calculating contribution to 
the polarization); they have the same number of vertices; they have 
the same kinds of lines connecting the vertices (continuous lines); 
and they have the same initial source, i. e., the temperature fluctuations (the 
dashed lines). Hence, it is better to write eq. \eqref{pol_sec_ord_diag} in 
terms of the diagram on the left-hand side of eq. \eqref{equiv_diag}, 
since it is easier to draw these types of diagrams when there is a large 
number of scatterings, as shown in eq. \eqref{a_lm_tot_diag}. 

In general, to order $n$:
\begin{equation}
\bar{\pi}_{lm}^{(n)} = \sum_{q=0}^{n-1} \pi_{lmq}^{(n)} 
(\Delta \eta_0, \Delta \eta_1, \ldots, \Delta \eta_n, X) \, 
F_{(l+q)2} (\Delta \eta_0, X; \Delta \eta_1, \ldots, \Delta \eta_n) \, .
\end{equation}
Explicit formulae for all the coefficients $\pi_{lmq}^{(n)}$ are not known explicitly, 
but can be computed iteratively after the identification of the diagrams that contribute 
to some order, and by employing the techniques presented in section \ref{high_power}. 
Naturally, the computation of $\bar{\pi}_{lm}^{(n)}$ requires the knowledge of 
$\bar{\theta}_{lm}^{(n-1)}$, the computation of $\bar{\theta}_{lm}^{(n-1)}$ 
requires the knowledge of both $\bar{\theta}_{lm}^{(n-2)}$ and 
$\bar{\pi}_{lm}^{(n-2)}$, and so on and so forth
--0 therefore, in this sense these coefficients form a hierarchy.

\subsection{Diagrammatic rules}

If we accept that Boltzmann's equations, as presented in 
eqs. \eqref{temperatura}, \eqref{temperatura_coef}, \eqref{polarizacao} and 
\eqref{polarizacao_coef}, are accurate enough to allow the computation of the
CMB temperature and polarization to all orders, then we can ask 
how the diagrams for our series expansion should be built. 
Clearly, when further precision is required, quadratic and even higher-order 
corrections for the metric perturbations may have to be included in 
Boltzmann's equation, and at some point the diagrams presented here may 
no longer improve the accuracy of the result.
However, if we assume that the Einstein-Boltzmann system of equations is exact, 
then all contributions can be written as a sequence diagrams, whose building 
blocks are presented in table~\ref{table}.

%%%%%%%%%%%%%%%%%%%%%%%%%%%%%%%%%
\begin{fmffile}{diag}
\begin{table}
\centering
\begin{tabular}{ l c r }
\hline
diag. segment  $\phantom{space}$   & numerical factor   $\phantom{space}$  & Bessel function \\
\hline \\
\parbox{20mm}{\begin{fmfgraph}(50, 60)
\fmfstraight
\fmfpen{thin}
\fmfright{i1,i2,i3}
\fmfleft{o1,o2,o3}
\fmf{phantom}{o1,o2,o3}
\fmf{vanilla}{i1,i2,i3}
\fmf{dashes}{i2,o2}
\fmfv{decor.shape=circle,decor.filled=empty,decor.size=.1w}{i2}
\end{fmfgraph}} & $-\frac{3}{2\pi} \sqrt{\frac{(l+2)!}{(l-2)!}}$  & $ \frac{j_l(k \Delta)}{(k \Delta)^2}$ \\
 &   &    \\
\parbox{20mm}{\begin{fmfgraph}(50, 60)
\fmfstraight
\fmfpen{thin}
\fmfright{i1,i2,i3}
\fmfleft{o1,o2,o3}
\fmf{phantom}{o1,o2,o3}
\fmf{dashes}{i1,i2,i3}
\fmf{dashes}{i2,o2}
\fmfv{decor.shape=circle,decor.filled=empty,decor.size=.1w}{i2}
\end{fmfgraph}} &  $\frac{1}{2}$ &  $ \left[1+3 \frac{\partial^2}{\partial(k\Delta)^2} \right] j_l(k \Delta)$ \\
 &   &   \\
\parbox{20mm}{\begin{fmfgraph}(50, 60)
\fmfstraight
\fmfpen{thin}
\fmfright{i1,i2,i3}
\fmfleft{o1,o2,o3}
\fmf{phantom}{o1,o2,o3}
\fmf{dashes}{i1,i2,i3}
\fmf{vanilla}{i2,o2}
\fmfv{decor.shape=circle,decor.filled=empty,decor.size=.1w}{i2}
\end{fmfgraph}} & $1$ &  $\frac{j_2(k \Delta)}{(k \Delta)^2}$ \\
 &   &   \\
\parbox{20mm}{\begin{fmfgraph}(50, 60)
\fmfstraight
\fmfpen{thin}
\fmfright{i1,i2,i3}
\fmfleft{o1,o2,o3}
\fmf{phantom}{o1,o2,o3}
\fmf{vanilla}{i1,i2,i3}
\fmf{vanilla}{i2,o2}
\fmfv{decor.shape=circle,decor.filled=empty,decor.size=.1w}{i2}
\end{fmfgraph}} & $1$ &  $\frac{j_2(k \Delta)}{(k \Delta)^2}$ \\
 &   &   \\
\parbox{20mm}{\begin{fmfgraph}(50, 60)
\fmfstraight
\fmfpen{thin}
\fmfright{i1,i2,i3}
\fmfleft{o1,o2,o3}
\fmfbottom{b1,b2,b3}
\fmftop{t1,t2,t3}
\fmf{phantom}{o1,o2,o3}
\fmf{phantom}{i1,i2,i3}
\fmf{phantom}{t2,v}
\fmf{dashes}{v,b2}
\fmf{dashes}{i2,o2}
\fmfv{decor.shape=circle,decor.filled=empty,decor.size=.1w}{i2}
\fmfv{decor.shape=circle,decor.filled=empty,decor.size=.1w}{v}
\end{fmfgraph}} &  $\frac{1}{2}$  &  $\left[1+3 \frac{\partial^2}{\partial(k\Delta)^2} \right] j_2(k \Delta)$ \\
\parbox{20mm}{\begin{fmfgraph}(50, 60)
\fmfstraight
\fmfpen{thin}
\fmfright{i1,i2,i3}
\fmfleft{o1,o2,o3}
\fmfbottom{b1,b2,b3}
\fmftop{t1,t2,t3}
\fmf{phantom}{o1,o2,o3}
\fmf{phantom}{i1,i2,i3}
\fmf{phantom}{t2,v}
\fmf{dashes}{v,b2}
\fmf{vanilla}{i2,o2}
\fmfv{decor.shape=circle,decor.filled=empty,decor.size=.1w}{i2}
\fmfv{decor.shape=circle,decor.filled=empty,decor.size=.1w}{v}
\end{fmfgraph}} &  $9$ &  $\frac{j_2(k \Delta)}{(k \Delta)^2}$ \\
\hline
\end{tabular}
\caption{\label{table} Building blocks for general diagramms.}
\end{table}
\end{fmffile}

As a general rule of thumb, the functions $j_l(kX)$ and $j_2(k\Delta\eta_n)$ 
must appear as the first and last elements inside the $k$ integral. 
Here, $\eta_n$ stands for the last index in the partition of the time 
interval $(\eta_o-\eta)$. The intervals $\Delta$ must be replaced by 
the convenient difference of consecutive conformal times which represent 
the initial and the final instants of that segment of the random flight. 
Outside the $k$ integral there should be the integral over the spatial
position of the primary source,
$\int dX \, X^2 \, S_{lm}(X, \eta)$. Finally, we have the integrals 
over conformal time weighted by visibility functions --- except the integral 
over $\eta$, which does not carry a factor of the visibility function.

For example, the following diagram represents the third-order contribution 
to polarization that arises from the polarization at second-order, when
that second-order polarization is due to the first-order correction to the temperature:
\begin{fmffile}{examp}
\begin{eqnarray}
\parbox{20mm}{\begin{fmfgraph}(50, 70)
\fmfstraight
\fmfpen{thin}
\fmfright{i1,i2,i3,i4,i5}
\fmfleft{o1,o2,o3,o4,o5}
\fmfbottom{b1,b2,b3,b4,b5}
\fmftop{t1,t2,t3,t4,b5}
\fmf{phantom}{o1,o2,o3,o4,o5}
\fmf{vanilla}{i1,i2,i3,i4,i5}
\fmf{phantom}{t3,v2,v3,b3}
\fmf{dashes}{v2,v3,b3}
\fmf{dashes}{o2,v3}
\fmf{phantom}{v3,i2}
\fmf{vanilla}{i3,v2,o3}
\fmfv{decor.shape=circle,decor.filled=empty,decor.size=.1w}{i3}
\fmfv{decor.shape=circle,decor.filled=empty,decor.size=.1w}{v3}
\fmfv{decor.shape=circle,decor.filled=empty,decor.size=.1w}{v2}
\end{fmfgraph}}
& = &    -  \frac{9}{2}  \, \frac{3}{2 \pi} \, \sqrt{\frac{(l+2)!}{(l-2)!}}  
\int_0^{\eta_o} d \eta_1 \, g(\eta_1) 
\int_0^{\eta_1} d \eta_2 \, g(\eta_2) 
\int_0^{\eta_2}  d \eta_3 \, g(\eta_3) 
\int_0^{\eta_3} d \eta 
\nonumber \\ 
& & \times \int dX \, X^2 \, S_{lm} (X, \eta) 
\Bigg\{
\int dk \, k^2 \, j_l(kX) \, \frac{j_l(k\Delta \eta_0)}{(k\Delta \eta_0)^2} 
\, \frac{j_2(k\Delta\eta_1)}{(k\Delta\eta_1)^2} 
\nonumber \\ 
& & \times 
\left[ 1 + 3 \frac{\partial^2}{\partial(k\Delta\eta_2)^2} \right]
 j_2(k\Delta\eta_2) \, j_2(k\Delta\eta_2) 
 \Bigg\} \, ,
\end{eqnarray}

The diagram below, on the other hand, represents a third-order contribution to
polarization arising from the second-order correction to the temperature,
when that second-order correction itself comes from the first-order 
correction to the temperature:
\begin{eqnarray}
\parbox{20mm}{\begin{fmfgraph}(50, 70)
\fmfstraight
\fmfpen{thin}
\fmfright{i1,i2,i3,i4,i5}
\fmfleft{o1,o2,o3,o4,o5}
\fmfbottom{b1,b2,b3,b4}
\fmftop{t1,t2,t3,t4}
\fmf{phantom}{o1,o2,o3,o4,o5}
\fmf{vanilla}{i1,i2,i3,i4,i5}
\fmf{phantom}{t2,v1}
\fmf{dashes}{v1,b2}
\fmf{phantom}{t3,v2}
\fmf{dashes}{v2,b3}
\fmf{dashes}{i3,o3}
\fmfv{decor.shape=circle,decor.filled=empty,decor.size=.1w}{i3}
\fmfv{decor.shape=circle,decor.filled=empty,decor.size=.1w}{v1}
\fmfv{decor.shape=circle,decor.filled=empty,decor.size=.1w}{v2}
\end{fmfgraph}}
& = &    -  \frac{1}{2} \, \frac{1}{2}  \, 
\frac{3}{2 \pi} \, \sqrt{\frac{(l+2)!}{(l-2)!}}  
\int_0^{\eta_o} d \eta_1 \, g(\eta_1)
\int_0^{\eta_1} d \eta_2 \, g(\eta_2) 
\int_0^{\eta_2} d \eta_3 \, g(\eta_3) 
\int_0^{\eta_3} d \eta 
\nonumber\\ & & 
\times 
\int dX \, X^2 \, S_{lm} (X, \eta) \,
\Bigg\{
\int dk \, k^2 \, j_l(kX) \, 
\frac{j_l(k\Delta \eta_0)^2}{(k\Delta \eta_0)}
\nonumber\\ & & \times
\left[ 1 + 3 \frac{\partial^2}{\partial(k\Delta\eta_1)^2} \right] 
j_2(k\Delta\eta_1) 
\left[ 1 + 3 \frac{\partial^2}{\partial(k\Delta\eta_2)^2} \right]
 j_2(k\Delta\eta_2) \, j_2(k\Delta\eta_3) 
 \Bigg\} \, .
\end{eqnarray}
\end{fmffile}
%%%%%%%%%%%%%%%%%%%%%%%%%%%%%%%%%

\section{Fourier-Bessel series for the extended random flight probabilities}
\label{fourier_bessel_sec}
From the previous arguments it is obvious that a full description of 
CMB temperature and polarization depends upon the evaluation of 
the extended random flight integrals $F_{l2}$. 
As we showed in eq. \eqref{pol_prob}, the extended random flights can be 
expressed as marginalizations over random flight probability densities. 
Unfortunately, explicit formulae for random flight probability densities in seven 
dimensions are not known, which means that at orders higher than three
the computations start to become highly complex. As a computational 
tool to deal with the extended random flight integral, we will introduce 
their series expansion in terms of a  Fourier-Bessel series 
--- this was also suggested by \cite{merzbacher}.

We will expand the function $F_{l L}(R, X; r_1, \ldots, r_n)$,
which is ubiquitous in our previous discussions (just substitute $\Delta\eta_0 \rightarrow R$,
$\Delta\eta_1 \rightarrow r_1$, etc.), and which was 
defined in eq. \eqref{F_lm_m}, in terms of a Fourier-Bessel series.
Given the behavior of this function and its interpretation, the segments
$R, r_1, \ldots, r_n$ are fixed for any particular flight. 
Since $r=\sqrt{R^2 + X^2 -2 \, R  \, X  \, \cos \alpha} \leq r_1 + r_2 + \ldots + r_n =: S_n$, it follows that the variable $X$ must have itself an upper limit
--- see figure 1 for the geometry involved in these limits. 
In fact, in the limit where all flight steps $r_1, \ldots, r_n$ are collinear, we have that 
$r=S_n$, and in that case $X$, $R$ and $S_n$ form the sides of a triangle.
Hence, the maximum value of $X$ is determined jointly by both $R$ and $S_n$, as
$0 \leq X \leq R + r_1 +r_2 + \ldots + r_n :=\tilde{S}_n$.

Now, we know that well-behaved functions with a limited domain can 
be expressed in terms of a Fourier-Bessel series \cite{hochstadt}. 
In fact, since $F_{lL}$ is one such well-behaved function, we have:
\begin{equation}
F_{lL}(R,X ; r_1, \ldots, r_n) = 
\sum_{i=1}^{\infty} f_{l L i}^{(n)} \, j_l \left( x_{li}^{(n)} \,X \right) \; ,
\end{equation}
where $x_{li}^{(n)}=z^l_i/\tilde{S}_n$, and $z^l_i$ is the $i$-th root of the
spherical Bessel function $j_l(z)$.
The coefficients $f_{l L i}^{(n)}$ are given by

\begin{equation}
\label{bessel_coef}
f_{lL i}^{(n)} = \frac{2}{\tilde{S}_n^3 \, j_{l+1}^2 (z^l_i )} 
\int_0^{\tilde{S}_n} dX \, X^2 \, j_l \left(  x_{li}^{(n)} \, X \right) \,
F_{lL}(R, X; r_1, \ldots, r_n) \; .
\end{equation}
Using eq. \eqref{pol_prob} we obtain:
\begin{equation}
\label{FX_calculo}
F_{lL}(R,X; r_1, \ldots, r_n) = \sum_{i=1}^{\infty} 
\frac{\pi}{\tilde{S}_n^3 \,  j_{l+1}^2(z^l_i)}
 \,  \frac{j_l \left( x_{li}^{(n)} \, R \right)}{\left( x_{li}^{(n)} \, R \right)^L}  \,  
 j_l \left( x_{li}^{(n)} \, X \right) 
 \, \prod_{q=1}^{n-1} \frac{j_L \left( x_{li}^{(n)} \, r_q \right)}{\left( x_{li}^{(n)} \, r_q \right)^L}  \,  
j_L \left( x_{li}^{(n)} \, r_n \right) \; .
\end{equation}
Now, eq. \eqref{produto_bessel} allows us to write \eqref{FX_calculo} in the form:
\begin{eqnarray}
\label{FX_int_dois}
F_{lL}( R, X; r_1, \ldots, r_n) & = & 
\frac{(-1)^L}{2} \frac{\pi}{\tilde{S}_n^3} 
\int_0^{\tilde{S}_n} d r   \, 
\left( \frac{R X}{r} \right)^{L-1} 
\frac{1}{R^L} \, P_l^{-L}(\cos \alpha)   \, 
(\sin \alpha)^L 
\nonumber \\ 
& & \times 
\sum_{i=1}^{\infty} \frac{1}{ j_{l+1}^2(z^l_i)} 
 \, j_m \left( x_{li}^{(n)} r \right) 
\prod_{q=1}^{n-1} \frac{j_L \left( x_{li}^{(n)}  \, r_q \right)}{\left( x_{li}^{(n)}  \,  r_q \right)^L}   \, 
j_L \left( x_{li}^{(n)}  \, r_n \right) \; , 
\end{eqnarray}
where $r^2 = R^2 + X^2 -2 \, R \, X \,  \cos\alpha$.

The extended random flight integrals can, therefore, be 
much more easily computed with the help of eq. \eqref{FX_int_dois}, 
since in that representation only a discrete number of modes need to be added, 
instead of the full-fledged integral that defines the random flight probability density.

%%%%%%%%%%%%%%%%%%%%%%%%%%%%%%%%%%%%%%%%%%%%
\section{Large number of scatterings and the H-theorem}
\label{H_theorem_sec}
In this Section we shall address a more fundamental question. Up to now we 
have presented a formalism that allows us to express the CMB polarization and 
temperature in position space, in terms of an expansion over the number of 
interactions the photons have suffered during recombination. 
However, at this point we can ask: up to what order the contributions should to be taken 
into account in order to yield an accurate expression for the physical observables?
But this question can only be answered if a concrete problem is given, 
so there is no general answer. Hence, we will invert this question, and ask: 
how important is it, for the observables, that we reach a certain order $n$ in the 
expansion? To be more specific, we will study the contributions arising from very 
high $n$ terms in eq. \eqref{a_lm_total}.

Since all computations depend on the probability densities for random flights
with $n$ steps, we will analyze the behavior of 
$p_n(r; \Delta \eta_1, \ldots , \Delta \eta_n \, | \, 7)$, for large $n$. 
Although explicit formulae for these probability functions are not known, 
an asymptotic expression is known in the case where all steps have the
same lengths, $\Delta \eta_1 = \Delta \eta_2 = \ldots= \Delta \eta_n =: \Delta$, 
and $n \to \infty$ \cite{watson}. In that case we have:
\begin{eqnarray}
p_n(r; \Delta, \ldots, \Delta \, | \, 7) 
& = & \frac{1}{\Gamma(7/2)} \left( \frac{7 r^2}{2n\Delta^2} \right)^{5/2}
\left( \frac{7 r}{2n\Delta^2} \right) 
\nonumber \\ 
& & \times 
\Bigg[ \phantom{}_1F_1 \left( \frac{7}{2}; \frac{9}{2}; \frac{-7 r^2}{2n\Delta^2} \right) 
-
\left( \frac{7 r^2}{9 n\Delta^2} \right) \,
\phantom{}_1F_1 \left( \frac{9}{2}; \frac{11}{2}; \frac{-7 r^2}{2n\Delta^2} \right) 
\Bigg] \; ,
\end{eqnarray}
where: 
\begin{displaymath}
\phantom{}_1F_1(a; c; z) = \frac{\Gamma(c)}{\Gamma(a)}
\sum_{k=0}^{\infty} \frac{\Gamma(a +k)}{\Gamma(c + k)} \frac{z^k}{k!}
\end{displaymath}
is the confluent hypergeometric function \cite{hochstadt}. 
Expanding $\phantom{}_1F_1$ in a power series we obtain:
\begin{eqnarray}
p_n(r; \Delta, \ldots, \Delta \, | \, 7) & = & \frac{7/2}{\Gamma(7/2)}  \left( \frac{7 r^2}{2n\Delta^2} \right)^{5/2} \left( \frac{7 r}{2n\Delta^2} \right) \nonumber\\ & & \times 
\sum_{k=0}^{\infty} \frac{1}{k!} \frac{ \left( \frac{9}{2} - \frac{7 r^2}{2n\Delta^2} \right) +
k \left( 1 - \frac{r^2}{n\Delta^2} \right)}{\left( \frac{7}{2} +k \right) \left( \frac{9}{2} + k \right)}
\left(\frac{-7 r^2}{2n\Delta^2}\right)^k \; .
\end{eqnarray}
Now, consider the sum:
\begin{displaymath}
\sum_{k=0}^{\infty} \frac{1}{k!} \frac{ \left( \frac{9}{2} -z \right) +
k \left( 1 - \frac{2}{7}z \right)}{\left( \frac{7}{2} +k \right) \left( \frac{9}{2} + k \right)}
\left(-z \right)^k 
 \leq  \sum_{k=0}^{\infty} \frac{1}{k!} \frac{1}{\left( \frac{7}{2} +k \right)} \left(-z \right)^k 
\leq \mathrm{e}^{-z} \; .
\end{displaymath}
It follows, therefore, that:
\begin{equation}
p_n(r; \Delta, \ldots, \Delta \, | \, 7) \leq  \frac{1}{r} \left( \frac{7 r^2}{2n\Delta^2} \right)^{7/2}
\mathrm{e}^{- \frac{7 r^2}{2n\Delta^2} } \, .
\end{equation}

Consequently, for fixed $r > 0$, $\Delta>0$, and $r \leq n \Delta$,
\begin{displaymath}
\lim_{n \to \infty} p_n(r; \Delta, \ldots, \Delta \, | \, 7) = 0 \, .
\end{displaymath}
Since $p_n$ is a normalized probability density, 
we conclude that, in the limit above, the probability distribution 
collapses into a Dirac delta-function centered at the origin, 
which is in agreement with the general features attributed to 
the random flight probability densities \cite{franceschetti}. 
In other words, increasing the number of changes of directions (i.e., scatterings) 
in a random flight leads to trajectories with increasingly smaller displacements 
from the origin. 

Let's investigate the consequences of this fact for the polarization 
through eq. \eqref{a_lm_prob}. Consider, in that respect, the integral:
\begin{eqnarray}
{\cal{S}}_{lm}
&=& \lim_{n \to \infty} \int_0^{\eta_o - \eta} dX  X^2  
S_{lm}(X, \eta)  \int d (\cos \alpha) 
\left( \frac{X \Delta}{r^3}\right)^2  P_l^{-2}(\cos \alpha)  (\sin \alpha)^2 \, 
p_n(r; \Delta, \ldots, \Delta  |  7) 
\nonumber \\ 
& = &
\int_0^{\eta_o - \eta} dX \, X^2  \, 
S_{lm}(X, \eta)  \int dr \, 
\frac{r}{X \Delta \eta_0} \, 
\left( \frac{X \Delta}{r^3} \right)^2 
\, P_l^{-2}(\cos \alpha) \, (\sin \alpha)^2  \, \delta(r) 
\nonumber \\ & = &
\int_0^{\eta_o - \eta} dX \, X^2 \,  
S_{lm}(X, \eta)  \, \frac{\Delta^2}{(\Delta \eta_0)^2} 
\lim_{r \to 0} \left[ \frac{X \Delta \eta_0}{r^5} \, P_l^{-2}(\cos \alpha) \, (\sin \alpha)^2 \right] \; .
\end{eqnarray}
We know, however, from either \cite{watson} or \cite{cmb_box}, that:
\begin{equation}
\int_0^{\infty} dk \, j_l(kr_1) \, j_l(kr_2) \, j_2(kr_3)  = \frac{\pi}{4} \, 
\frac{r_1 r_2}{r_3^3} \,
P_l^{-2}(\cos \alpha) \, (\sin \alpha)^2 \; ,
\end{equation}
and also that \cite{cmb_box}:
\begin{equation}
\lim_{r_3 \to 0} \int_0^{\infty} dk \, j_l(kr_1) \, j_l(kr_2) \, j_2(kr_3)  
= \frac{r_2^2}{r_1^4} \, \frac{r_3^2}{15} \, 
\frac{\pi}{2} \delta (r_1 - r_2) \, .
\end{equation}
Therefore, we have that:
\begin{eqnarray}
{\cal{S}}_{lm} (\eta_0,\eta)
& = & \int_0^{\eta_o - \eta} dX \, X^2 \,  
S_{lm}(X, \eta)  \, \frac{\pi}{30} \, \frac{\Delta^2}{X^4} 
\delta( \Delta \eta_0 - X) 
\nonumber \\ 
& = & 
\frac{\pi}{30} \, \frac{\Delta^2}{(\Delta \eta_0)^2} \,
S_{lm}(\Delta \eta_0, \eta) \, .
\end{eqnarray}

The final lesson is that, in the limit of an infinite number of scatterings, 
we basically end with instantaneous recombination --- with the crucial
caveat that the source term is damped by $(\Delta/\Delta \eta_0)^2 \to 0$.
The conclusion is, therefore, that no net polarization is generated at the limit 
of infinite number of scatterings. Similarly, for the temperature fluctuations the 
derivatives of the integral which appears in $\theta_{lm}^{(n)}$ ensure that 
the signal vanishes as well.

The large $n$ limit, therefore, cannot contribute significantly to the final 
observables. If we regard this order as the number of scatterings that 
a photon has experienced during recombination, then the collapse of 
the probability density associated with random flights expresses a 
fundamental physical fact: increasing the number of scatterings takes 
the system closer to thermodynamical equilibrium. As we know, in 
thermodynamical equilibrium the CMB temperature is given simply 
by the Planck distribution, with no distortions (apart from the
dipole which is induced by the velocity of baryons). 
Maximizing the von Neumann entropy also ensures that, in 
equilibrium, no net polarization is generated. These two facts, together, 
constitute the expression of Boltzmann's H-theorem to the CMB 
\cite{ehlers}. 

Describing the CMB by means of random flight probability density functions 
provides, therefore, an illustration of the statement of the H-theorem: when 
few scatterings take place during recombination (low $n$), photons and 
electrons are surely out of equilibrium, and each interaction generates 
temperature fluctuations and polarization that are not sufficiently erased by 
the subsequent scatterings. As a consequence, the low-$n$ terms in the 
expansions account for the largest contributions for the signal of the
physical observables. For a large number os scatterings, the signal is 
washed out by further scatterings, which can be expressed in terms of 
the collapse of the probability density in the large-$n$ limit.

All that is left to discuss is the physical grounds for assuming (as we did 
in this Section) that all the steps performed in the random flights have the 
same length. Despite the fact that this working hypothesis was not proposed 
because of any physical reason, but purely because of mathematical convenience, 
we can nevertheless argue that, when the electron-photon system was 
close to the tight-coupling regime, the assumption that the visibility 
function is step-shaped is not completely crude.
Continuity with respect to the arguments of the probability density distribution 
makes our working hypothesis less unnatural.

\section{Conclusions}

The Boltzmann hierarchy for the problem of the evolution of temperature and 
polarization fluctuations of the CMB is equivalent to a system of coupled 
integral equations. We showed that this system can be written in position 
space, and that the objects that appear in this description are the same
probability density functions that appear in random flight problems. 
The emergence of random flight probability functions from Boltzmann's 
equations clarifies the physical interpretation that CMB temperature fluctuations 
and polarization are generated from scatterings of photons by low-energy 
electrons during the recombination. More importantly, we showed that the 
{\em number} of scatterings during recombination is a key ingredient in 
the description of this problem --- in fact, a perturbative expansion can be 
performed in terms of the number of scatterings. 

We showed, using asymptotic formulae for the random flight probability 
distribution, that contributions coming from high-order terms (i.e., many scatterings) 
should indeed be negligible. Since high-order terms represent the past history of 
photons that have scattered many times during recombination, we concluded
that their vanishing contribution to the CMB signal is an illustration of 
Boltzmann's H-theorem. In the context of the CMB, this theorem states that, in 
thermodynamical equilibrium, the CMB temperature should be given by a 
Planck spectrum, and that the net polarization should vanish.

Using the fact that the random flight distribution functions vanish identically 
if one cannot form a closed polygon from its intermediate steps, we presented 
Fourier-Bessel series expansions for the associated probability distribution 
function. These expressions lead to simple numerical recipes for the computation
of these distributions, since explicit formulae for them are not presently known for 
any dimension. Another very important result which derives from the emergence 
of random flights is the fact that, at each given time, the domains of dependence 
of the problem are compact sets. This is not obvious from the usual treatment of 
Boltzmann's equations in Fourier space.

From a more formal perspective, it would be important to understand 
which classes of Boltzmann's equations are amenable to a treatment 
in terms of random flight probability densities. Many of those problems
could be then examined in light of the formalism that has been 
developed for the study of the CMB.

Finally, we can foresee some possible applications of this work.
The series expansion in terms of the number of scatterings can
be used for numerical simulations of constrained maps of temperature 
and polarization. Due to the general vanishing property of the 
probability density functions for the extended random flight if intermediate
displacements do not form a polygon, 
and the decreasing of the visibility function for $z >> 10^3$, 
we can in practice take all the sources to vanish outside of a 
sphere of radius $R$ sufficiently large, and calculate the temperature and polarization 
corrections using Fourier-Bessel expansions, as shown in 
ref. \cite{cmb_box}. In Fourier-Bessel basis only a discretized tower of
modes contribute to each observable at each multipole, and 
the computational advantages of this approach are described 
in ref. \cite{leistedt}. In what concerns the convergence of the iterative process, 
depending on the desired accuracy, application or the angular scale that one 
wishes to examine, it may be sufficient to consider only the first couple of 
scatterings of the photons, since going further in the expansion would bring 
only contributions from terms highly suppressed by powers of the visibility 
function. We should emphasize that in the limit of large number of interactions 
the signal will be doubly suppressed: the first suppression comes from the high
order power of the visibility function, and the second from the vanishing
of the temperature and polarization corrections due to the collapse of 
the random flight probability density function.
Also corrections of higher orders could be comparable to the 
corrections coming from second order perturbations in 
space-time metric, or more detailed models for the
dynamics of interactions during the recombination.
The partial summation of some classes of diagrams (in the spirit of a 
renormalization group) can also be performed, which would lead to 
more accurate results without increasing the complexity.
In particular, the position-space analysis is especially suited for 
introducing gravitational lensing (work in progress).

%%%%%%%%%%%%%%%%%%%%%%%%%%%%%%%%%%%%%%%%%%%%
\appendix

%%%%%%%%%%%%%%%%%%%%%%%%%%%%%%%%%%%%%%%%%%%%
\section{Extended random flight integrals times even functions}

We shall show the possibility of expressing a larger set of integrals in term of random flight probability densities. Specifically, we will concentrate in expressions with the form:
\begin{equation}
\label{F_lm_m}
F_{lL}(R, X; r_1, \ldots, r_n) = \int_0^{\infty} dk \, k^2 \, j_l(kX) \, \frac{j_l(kR)}{(kR)^L} 
\prod_{q=1}^{n-1}\frac{j_L(k r_q)}{(kr_q)^L}  j_m(k r_n) \, .
\end{equation}
Our aim is to express integrals like:
\begin{equation}
\label{G_lL}
G_{lL}(R, X; r_1, \ldots, r_n) = \int_0^{\infty} dk \, k^2 \, f(k) \, j_l(kX) 
\, \frac{j_l(kR)}{(kR)^L} \prod_{q=1}^{n-1}\frac{j_L(k r_q)}{(kr_q)^L}  j_L(k r_n) \, 
\end{equation}
in terms of combinations of $F_{l'L'}(R, X; r_1, \ldots, r_n)$, 
for some adequate $l'$ and $L'$.
In $G_{lL}$, $f(k)$ is taken to be a real function such that its Taylor 
series carries only terms with even powers of $k$. 
The simplest such function is $a \, k^0$, for $a \in \bold{R}$. 
Naturally, in that case the connection we are looking for is quite obvious. 
We will, therefore, study monomials in the Taylor expansion, namely, we 
will take $f(k)=k^2$.

\subsection{The case of $G_{lL}(R, X; r_1, \ldots, r_n)$ when $f(k)=k^2$}
The following will be the object of our attention:
\begin{equation}
\label{G_lm_2}
G_{lL}^{(2)}(R, X; r_1, \ldots, r_n) = \int_0^{\infty} dk \, k^2 \, k^2 \, 
j_l(kX) \, \frac{j_l(kR)}{(kR)^L} \prod_{q=1}^{n-1} \frac{j_L(k r_q)}{(kr_q)^L}  j_L(k r_n) \; ,
\end{equation}
where the superscript indicates that we are treating the power $k^2$ 
in the expansion of some $f(k)$.

Recall the recurrence relations obeyed by spherical Bessel functions:
\begin{equation}
\label{bessel_esferica_recorrencia}
\frac{d}{dx} \left[ x^{\nu+1} j_{\nu}(x) \right] = x^{\nu+1} j_{\nu-1}(x) \; .
\end{equation} 
It then follows that, for some integer $l$, that: 
\begin{displaymath}
x^{l+2} j_l(x) = \frac{d}{dx}(x^{l+2} j_{l+1}(x)) \; .
\end{displaymath}
The chain rule gives:
\begin{displaymath}
\frac{d}{dx} \left\{ y(x)^{l+2} j_{l+1}[y(x)] \right\} =
\frac{d}{dy} \left\{ y(x)^{l+2} j_{l+1}[y(x)] \right\} \, \frac{dy}{dx} \; .
\end{displaymath}
If $y(x)=kx$, 
\begin{equation}
\label{jl_jl_plus_one}
j_l(kx) = \frac{1}{k} \frac{1}{(kx)^{l+2}} \frac{d}{dx} \left[ (kx)^{l+2} j_{l+1}(kx) \right] \, .
\end{equation}

Inserting \eqref{jl_jl_plus_one} into \eqref{G_lm_2} we obtain:
\begin{eqnarray}
G_{lL}^{(2)}(R, X; r_1, \ldots, r_n) 
& = & \int_0^{\infty} dk \, k^4 \, 
\frac{1}{(kR)^L} 
\frac{1}{k} 
\frac{1}{(kR)^{l+2}} 
\frac{d}{dR} \left[ (kR)^{l+2} j_{l+1}(kR) \right] 
\nonumber \\
& & \times  \frac{1}{k} \frac{1}{(kX)^{l+2}} \frac{d}{dX} \left[ (kX)^{l+2} j_{l+1}(kX) \right]
\prod_{q=1}^{n-1}\frac{j_L(k r_q)}{(kr_q)^L} 
j_L(k r_n) \nonumber\\
& = & \frac{1}{(RX)^{l+2}} \int_0^{\infty} dk \, k^2 \, \frac{1}{(kR)^L} 
\frac{d}{dR} \left[ R^{l+2} j_{l+1}(kR) \right] 
\nonumber \\ 
& & \times  
\frac{d}{dX} \left[ (kX)^{l+2} j_{l+1}(kX) \right]
\prod_{q=1}^{n-1}\frac{j_m(k r_q)}{(kr_q)^L} 
j_L(k r_n) \; . 
\nonumber
\end{eqnarray}
However,
\begin{displaymath}
\frac{1}{(kR)^L} \frac{d}{dR} u(R) = \frac{d}{dR} \left[\frac{u(R)}{(kR)^L} \right]
+ \frac{L}{R} \frac{u(R)}{(kR)^L} \, 
\end{displaymath}
for any function $u(R)$. Hence,
\begin{eqnarray}
G_{lL}^{(2)}(R, X; r_1, \ldots, r_n) & = & \frac{1}{(RX)^{l+2}} \frac{d}{dX} 
\Bigg\{
\frac{d}{dR} \Bigg[ (RX)^{l+2} \Bigg( \int_0^{\infty} dk \, k^2 \, \frac{j_{l+1}(kR)}{(kR)^L}
j_{l+1}(kX)  
\nonumber\\ & & \times 
\prod_{q=1}^{n-1}\frac{j_L(k r_q)}{(kr_q)^L} \, j_m(k r_n) \Bigg) \Bigg] 
+ \frac{m}{R} (RX)^{l+2} \nonumber\\ & & \times 
\Bigg( \int_0^{\infty} dk \, k^2 \, \frac{j_{l+1}(kR)}{(kR)^L}
\, j_{l+1}(kX) 
\prod_{q=1}^{n-1}\frac{j_L(k r_q)}{(kr_q)^L}  j_L(k r_n) \Bigg) \Bigg\} \, .
\end{eqnarray}

We obtained, therefore, a relation:
\begin{equation}
\label{G_lm_k2_resultado}
G_{lL}^{(2)}(R, X; r_1, \ldots, r_n) = \frac{1}{(RX)^{l+2}} \frac{\partial}{\partial X} 
\left( \frac{\partial}{\partial R}
+ \frac{L}{R} \right) \left[ (RX)^{l+2} F_{(l+1)L}(R, X; r_1, \ldots, r_n) \right] \; ,
\end{equation}
i. e., $G_{lL}^{(2)}(R, X; r_1, \ldots, r_n)$ can be expressed in terms
of  $F_{(l+1)L}(R, X; r_1, \ldots, r_n)$, which is the result we 
were looking for.

\subsection{Higher powers of $f(k)$}
\label{high_power}
We will not derive a closed-form expression for 
$G_{lL}^{(q)}(R, X; r_1, \ldots, r_n)$ in the case of any even $q$, 
however, we will illustrate how to generalize the argument presented in 
the previous subsection to these higher powers.

First, we point out that higher powers in the expansion of $f(k)$ must be 
canceled out by corresponding factors $1/k$ when
eq. \eqref{jl_jl_plus_one} is iterated for the appropriate number of times. 
For the fourth degree monomial in the expansion of $f(k)$ we would have:
\begin{eqnarray}
j_l(kx) & = & \frac{1}{k} \frac{1}{x^{l+2}} \frac{d}{dx} \left[ x^{l+2} \left( 
\frac{1}{k} \frac{1}{x^{l+3}} \frac{d}{dx} \left[ x^{l+3} j_{l+2}(kx) \right] \right) \right]
\nonumber\\ & = & 
\frac{1}{k^2} \left[ \frac{1}{x^{l+3}} \frac{d^2}{dx^2} ( x^{l+3} j_{l+2}(kx)) 
- \frac{1}{x^{l+4}} \frac{d}{dx} (x^{l+3} j_{l+2}(kx)) \right] \, .
\end{eqnarray}
This holds for $j_l(kR)$, as well as for $j_l(kX)$. We must bear in 
mind that, in the original equations, terms like $\frac{j_l(kR)}{(kR)^L}$ appear, 
and therefore we must change the orders of the derivatives with respect to 
$R$ with a term like $(kR)^{-L}$. This can be performed easily with the aid 
of the Leibniz formula for the $n$-th derivative of products of functions:
\begin{eqnarray} 
\frac{1}{(kR)^L} \frac{d^q}{dR^q}[u(R)]  & = &
\frac{d^q}{dR^q} \left( \frac{u(R)}{(kR)^L} \right) \nonumber\\ & & -  \sum_{r=0}^{q-1} 
\left( \begin{array}{c} q-1 \\ r \end{array} \right) \frac{d^r}{dR^r} [u(R)] \frac{(-1)^{q-r}}{(kR)^L}
\frac{L(L-1)\ldots(L-q-r)}{R^{q-r}} \; . 
\nonumber
\end{eqnarray}
This process must be iterated until all derivatives with respect to $R$ 
have been interchanged with terms like $(kR)^{-L}$.

As a final remark, we note that, even if the analytical problem is 
quite cumbersome, $G_{lL}^{(q)}(R, X; r_1, \ldots, r_n)$ can always 
be obtained, for even $q$, in terms of derivatives of the function $F_{(l+q/2) L}(R, X; r_1, \ldots, r_n)$.

%%%%%%%%%%%%%%%%%%%%%%%%%%%%%%%%%%%%
\acknowledgments
%%%%%%%%%%%%%%%%%%%%%%%%%%%%%%%%%%%%
The authors would like to thank Walter Wreszinski for useful comments.
This work was supported by FAPESP, and P.R. acknowledges the support of
the European Commission under the contract PITN-GA-2009-237920.

\bibliographystyle{plain}
\bibliography{referencias}

\end{document}